\documentclass[fleqn,usenatbib]{mnras}

\usepackage{newtxtext,newtxmath}

\usepackage[T1]{fontenc}
\usepackage[english]{babel}

\usepackage{csquotes} 

\DeclareRobustCommand{\VAN}[3]{#2}
\let\VANthebibliography\thebibliography
\def\thebibliography{\DeclareRobustCommand{\VAN}[3]{##3}\VANthebibliography}

\usepackage{graphicx}   
\usepackage{amsmath}    

\hypersetup{
  colorlinks=true, allcolors=blue
  }
\usepackage{xcolor}

\usepackage{booktabs}
\usepackage{subcaption}
\usepackage{multirow}

\usepackage[noabbrev]{cleveref}
\crefname{equation}{Eq.}{Eqs.}
\crefname{section}{Section}{Sections}
\crefname{figure}{Figure}{Figures}
\crefname{table}{Table}{Tables}
\crefname{appendix}{Appendix}{Appendices}
\Crefname{figure}{Figure}{Figures}
\Crefname{equation}{Equation}{Equations}
\Crefname{section}{Section}{Sections}
\Crefname{table}{Table}{Tables}

\newcommand{\degsq}[0]{~$\rm{deg^{2}}$}
\newcommand{\mdegsq}[0]{~$\rm{deg^{-2}}$}
\newcommand\ltapprox{\,\lower.6ex\hbox{$\buildrel <\over \sim$}\,}

\title[4MOST-CRS: BG and LRG Target Selection]{The 4MOST-Cosmology Redshift Survey: Target Selection of Bright Galaxies and Luminous Red Galaxies}
\author[Verdier, Rocher et al.]{
Aurélien Verdier,$^{1}$\thanks{aurelien.verdier@epfl.ch}
Antoine Rocher,$^{1}$\thanks{antoine.rocher@epfl.ch}
Behnood Bandi,$^{2}$
Johan Richard,$^{3}$
Boudewijn F. Roukema$^{4,3}$,\newauthor
Jon Loveday,$^{2}$
Elmo Tempel,$^{5}$
Maciej Bilicki,$^{6}$
Jean-Paul Kneib,$^{1}$
Mathilde Guitton$^{1}$
\\
$^{1}$Institute of Physics, Laboratory of Astrophysics, \'Ecole Polytechnique F\'ed\'erale de Lausanne (EPFL), Observatoire de Sauverny, CH-1290 Versoix, Switzerland\\
$^{2}$Astronomy Centre, University of Sussex, Falmer, Brighton BN1 9QH, UK \\
$^{3}$Univ Lyon, Univ Lyon1, ENS de Lyon, CNRS, Centre de Recherche Astrophysique de Lyon UMR5574, Saint-Genis-Laval, France \\
$^{4}$Institute of Astronomy, Faculty of Physics, Astronomy and Informatics, Nicolaus Copernicus University, Grudziadzka 5, 87-100 Toru\'n, Poland\\
$^{5}$Tartu Observatory, University of Tartu, Observatooriumi 1, 61602 Tõravere, Estonia \\
$^{6}$Center for Theoretical Physics, Polish Academy of Sciences, al. Lotników 32/46, 02-668 Warsaw, Poland
}

\date{Accepted XXX. Received YYY; in original form ZZZ}

\pubyear{\the\year{}}

\begin{document}
\label{firstpage}
\pagerange{\pageref{firstpage}--\pageref{lastpage}}
\maketitle

\begin{abstract}
The Cosmology Redshift Survey of the 4-metre Multi-Object Spectroscopic Telescope (4MOST-CRS) will provide redshift measurements of galaxies and quasars over 5700{\degsq} in the southern hemisphere. 
As targets for the 4MOST-CRS, we present a selection of an $r<19.25$ magnitude limited sample of Bright Galaxies (BG) and a colour selected sample of Luminous Red Galaxies (LRG) based on DESI Legacy Survey DR10.1 photometric data, in the redshift ranges $0.1<z<0.5$ and $0.4<z<1$, respectively.
These samples are selected using the $g$, $r$, $z$, and $W1$ (from unWISE) photometric bands. For BGs, the star--galaxy separation is performed based on Gaia and Tycho-2 star catalogues. Following 4MOST requirements, the target densities of BGs and LRGs are 250{\mdegsq} and 400{\mdegsq}, respectively. We quantified the stellar contamination to be $<2-4\%$ for both galaxy samples using data from the first DESI data release.
Using angular clustering, we show that both samples are robust against imaging systematics, confirming a low stellar contamination. Finally, we provide forecasts of the baryonic acoustic oscillation (BAO) and growth of structure ($\sigma_8$) measurements from the 4MOST-CRS alone and in combination with DESI. From the 4MOST-CRS, we expect 3\% and 25\% precision on BAO and $\sigma_8$ measurements, respectively. Combining 4MOST-CRS with DESI improves the constraints from DESI alone by 12$\%$ in the 0.4 < z < 0.8 redshift range, leading to the most stringent constraints on BAO measurements from spectroscopic galaxy clustering.
\end{abstract}




\section{Introduction}

In the mid-1980s, the cosmology community started presenting observational arguments in favour of a non-zero cosmological constant $\Lambda$ (\citealt{VittorioSilk1985,KofmanStarobinsky1985,Tayler1986,efstah}; for a recent review, see \citealt{Marcy2024DEprediscovery}). The interpretation of the observational constraints in terms of the “universe ... `accelerating’ {}” was introduced \citep{RoukemaYoshii1993} and strong evidence for non-zero $\Lambda$ (within the class of Friedmann--Lema\^{\i}tre--Robertson--Walker (FLRW) models) was found in the late ‘90s~\citep{Perlmutter1999, Riess1998}.
One of these probes, statistics of the large-scale structure (LSS) of the Universe, became especially useful: the baryon acoustic oscillations (BAO,~\citealt{Eisenstein_HU_1998}), which are imprinted in the comoving spatial distribution of galaxies.

Since the first detections of BAO in the Sloan Digital Sky Survey (SDSS,~\citealt{SDSS_2000}) and the Two-Degree Field Galaxy Redshift Survey~\citep{Cole2005}, large spectroscopic galaxy redshift surveys have probed the 3-dimensional distribution of galaxies in the past light cone, increasingly constraining the FLRW cosmological parameters. In addition to BAOs, the peculiar velocities of galaxies induce redshift space distortion (RSD,~\citealt{Kaiser1987}), which provides constraints on the growth of cosmic structures.
The SDSS experiment with the Baryon Oscillation Spectroscopic Survey (BOSS,~\citealt{SDSS2013}), ending in 2020, together with the extension eBOSS~\citep{eBOSS2021} provided measurements of $\sim 3.5$ million galaxy spectra, painting a clear picture of the $\Lambda$CDM standard cosmological model.

In 2021, the DESI experiment~\citep{DESI_overview2022} started observations to extend the concept of SDSS further, expecting to measure 40 million redshifts over $\sim14~000${\degsq} using bright galaxies (BG), luminous red galaxies (LRG), emission line galaxies (ELG), and quasars (QSO). 
After one year of observations, DESI observed more than $\sim10$ million galaxy spectra~\citep{DESI_DR1}, providing the most precise measurements so far of BAOs in the redshift range $0<z<2.1$~\citep{DESI_BAO_DR1}.
The first DESI data release (DR1) revealed a hint of evolving dark energy~\citep{DESI_KP72025}, which was strengthened with DESI DR2 (3 years of observations,~\citealt{DESI_BAO_DR2}).

In this context, the 4MOST-Cosmology Redshift Survey (4MOST-CRS) will be a major asset in confirming or rejecting these results by providing a totally independent measurement of redshifts in the southern hemisphere.
The 4-metre Multi-Object Spectroscopic Telescope (4MOST;~\citealp{4most_prop}) is a new high-multiplex wide-field spectroscopic instrument that will be installed on the Visible and Infrared Survey Telescope for Astronomy (VISTA;~\citealp{VISTA}) at the Paranal Observatory in Chile. 4MOST has 2436 fibres linked to high/low-resolution spectrographs (respectively 812 and 1624 fibres) over a wide field of view of 4.2 square degrees. The 4MOST consortium plans to conduct a 5-year survey targeting 18 different science cases (called surveys), covering a wide spectrum of astronomical themes from intra-galactic to cosmological scales. Each survey is itself divided into multiple experiments, called subsurveys.

Among these, the Cosmology Redshift Survey (CRS, survey number 8 (S08);~\cite{2019Msngr}), is designed to conduct the largest southern hemisphere spectroscopic galaxy redshift survey to date, over $\sim 5700 \ $ deg$^2$.
4MOST-CRS aims to perform spectroscopic clustering measurements of galaxies and quasars between redshifts 0.1 and 3.5, allowing cosmological parameters to be constrained through BAO and RSD measurements. Divided into four different subsurveys, the 4MOST-CRS will observe carefully selected samples of $\sim 1.4$ million of bright galaxies (S0801) at low redshift ($0.1<z<0.4$), $\sim 2.3$ million luminous red galaxies (S0802) at intermediate redshift ($0.4<z<1$),  $\sim 1.5$ million quasars (S0803) at high redshift ($0.9<z<2.2$) and up to $z<3.5$ for quasars-Ly$\alpha$ (S0804).
In addition, many key science questions will be addressed by combining galaxy and quasar spectroscopic samples from 4MOST-CRS with data from current or future photometric and radio facilities in the southern hemisphere such as the Vera Rubin Observatory (LSST) ~\citep{LSST}, the \textit{Euclid} mission~\citep{euclid}, the Hydrogen Intensity and Real-time Analysis eXperiment (HIRAX,~\cite{HIRAX}) and the Square Kilometre Array Observatory~\citep{SKA}.

This paper describes the selection of the 4MOST-CRS BG and LRG targets, assesses the uniformity of the selection, and the spectroscopic performance using data from the DESI Data Release~1 (DR1~\cite{DESI_DR1}).
The initial proposal described in~\cite{2019Msngr} has been modified in order to be more complementary to the DESI survey. We plan to use a similar selection of targets compared to DESI, allowing the easiest combination of galaxy samples between DESI and 4MOST-CRS. However, the unique observational strategy of 4MOST, designed to optimise multiple science objectives, requires the sharing of observational time among the 18 surveys. As a result, the target density of 4MOST-CRS tracers has been reduced compared to the DESI standard to align with the requirements of the 4MOST observational strategy.

We first describe the motivations and modifications of the CRS strategy compared to the first description of CRS in~\cite{2019Msngr}, namely the dropping of the emission line galaxies from the target samples, the switch from the VISTA imaging survey to the Legacy Survey (LS) DR10.1 and the reduction of the target density on the sky. In addition, we use the angular clustering of the target samples as a quality check of the target selection, investigating potential imaging systematics and mitigation techniques. An in-depth study of the clustering properties of the samples is performed in a companion study~\cite{Bandi_in_prep}. Finally, using a simulation of the 4MOST survey, we assess the expected completeness of the targets after 5 years of nominal observations and forecast the constraints on cosmological parameters of the 4MOST-CRS and the combination with the DESI survey.

The structure of the paper is as follows, \cref{sec:VISTA} describes the VISTA photometric data (\cref{sec:vista_phot}), the former CRS target selection (\cref{sec:vista_ts}) and the reasons why we changed to Legacy Survey DR10.1 photometric data (\cref{sec:vista_to_ls}).
The latter is described in \cref{sec:LS_target_sel} as well as both final target selections for BGs and LRGs (\cref{sec:LS_TS}). Furthermore, different tests to monitor the quality of the CRS selection are presented in \cref{sec:TS_Val} with a cross-correlation with DESI DR1 data (\cref{sec:DESI_DR1}), angular clustering of the samples (\cref{sec:AC_theorie}), the simulated completeness of the sample after the 5 years of 4MOST observation (\cref{sec:comp}) and cosmological forecasts of CRS samples compared and combined with DESI (\cref{sec:forecasts}).
Finally, \cref{sec:phot_sys} focuses on the mitigation of photometric systematics using linear regression and random forest techniques.

\section{Former target selection from VHS}
\label{sec:VISTA}

\subsection{The VISTA ZYJHKs photometry}
\label{sec:vista_phot}


\begin{table}
    \centering
    \caption{Area and depth reached in each band (5$\sigma$, AB) for VISTA (VHS + VIKING) and DESI Legacy Survey (DECaLS + DES) photometric surveys.}
    \begin{tabular}{ccc}
    \hline
    \hline
    Survey & Area & Depth\\
     & [deg$^2$] & [AB mag]\\
    \hline
    VHS-VST & \multirow{2}{*}{5000} & $Y=21.2$, $J=21.2$\\
    ATLAS & & $H=20.6$, $K_s=20.0$\\
    \hline
    VHS-DES & 4500 & $J=21.6$, $H=21.0$, $K_s=20.4$ \\
    \hline
    \multirow{2}{*}{VIKING} & \multirow{2}{*}{1350} & $J=22.1$, $H=21.5$, $K_s=21.2$ \\
     & & $Z=23.1$, $Y=22.3$ \\
    \hline
    \hline
    DECaLS & 9000 & $g=24.0$, $r=23.4$, $z=22.5$\\
    \hline
    \multirow{2}{*}{DES}& \multirow{2}{*}{5000} & $g=24.7$, $r=24.4$, $i=23.8$  \\
    & & $z=23.1$, $Y=21.7$\\
    \hline
    \end{tabular}

    \label{tab:phot_info}
\end{table}

The Visible and Infrared Survey Telescope for Astronomy (VISTA) is a 4-metre wide-field survey telescope at ESO’s Paranal Observatory. It has a 1.65 degree diameter field of view and is equipped with a near-infrared imaging camera (VIRCAM) that can operate at wavelengths 0.8 to 2.3\,$\mu$m.
For 5 years, the VISTA telescope performed six extensive photometric surveys with different depths and footprints in the sky in five different bands: $Z$, $Y$, $J$, $H$, and $K_s$ (\cref{tab:phot_info}).
The first target selection proposed by the 4MOST-CRS was based on the VISTA Hemisphere Survey (VHS, \citealt{vhs_dr5}) and the VISTA Kilo-Degree Infrared Galaxy Survey (VIKING, \citealt{viking_2013}). The bands used for the selection were $J$ and $K_s$. Combining VIKING and VHS allowed us to select targets across the entire southern hemisphere. The construction of the footprint was first motivated by the combination of DES, KiDS and VST-ATLAS that were not covered by DESI, reaching an area of 7500{\degsq}. To complete the selection, WISE photometric data were also used, in particular the $W1$ band (3.4 $\mu$m).
Hereafter, for brevity, we refer to the two VISTA surveys, VHS and VIKING, together as \enquote*{VISTA}.

\subsection{Initial target selection from VISTA photometry} \label{sec:vista_ts}




\begin{figure}
    \centering
    \includegraphics[width=\linewidth]{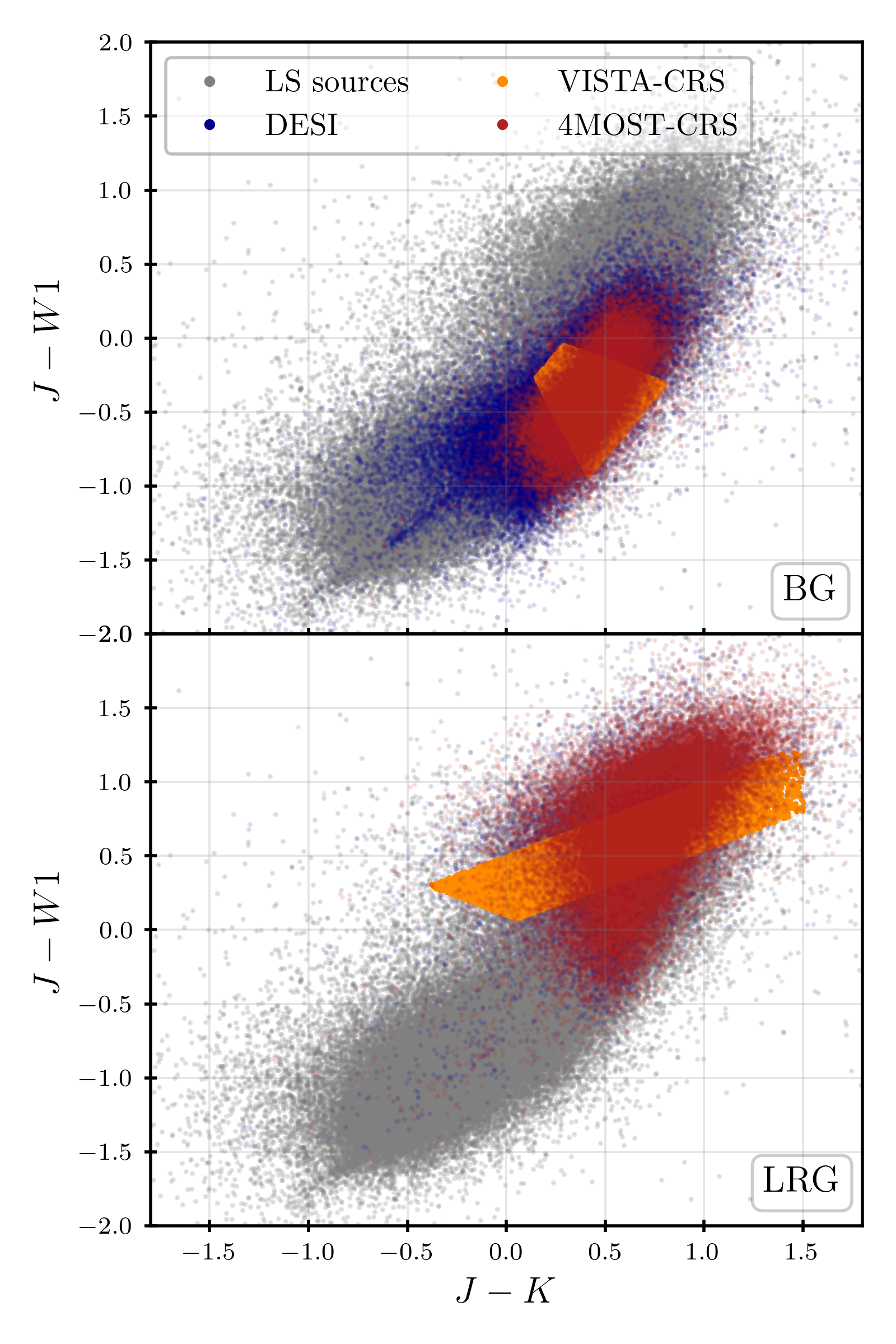}
    \vspace{-0.7cm}
    \caption{Colour-colour cut plots corresponding to the previous VISTA-CRS target selection using $J$, $K$ and $W1$. The area within which targets were chosen using VISTA is shown in orange. The red points correspond to the new targets selected by 4MOST-CRS using the modified DESI cuts. The dots in blue are the selected targets in the DESI survey, and in grey are all sources from the Legacy Survey. The upper panel shows the BGs and the lower panel the LRGs.}
    \label{fig:lrg_vista}
\end{figure}

In~\cite{2019Msngr}, BGs and LRGs were selected using the VISTA and WISE surveys.
The magnitudes selected to differentiate the different types of galaxies were $J$, $Ks$ and $W1$.
The Dark Energy Survey (DES) with $g$, $r$ and $i$ photometric bands were used to target ELGs. The ELGs sample has been dropped from the 4MOST-CRS program due to the large amount of fibre hours necessary to get enough completeness. 
In addition, \textit{Euclid} will cover a large part of the ELGs of the 4MOST-CRS footprint. The previous target selections for each galaxy sample are presented in \cref{tab:VISTA_cuts}, and colour cuts are shown in \cref{fig:lrg_vista} in orange.

\begin{figure}
    \centering
    \includegraphics[width=\linewidth]{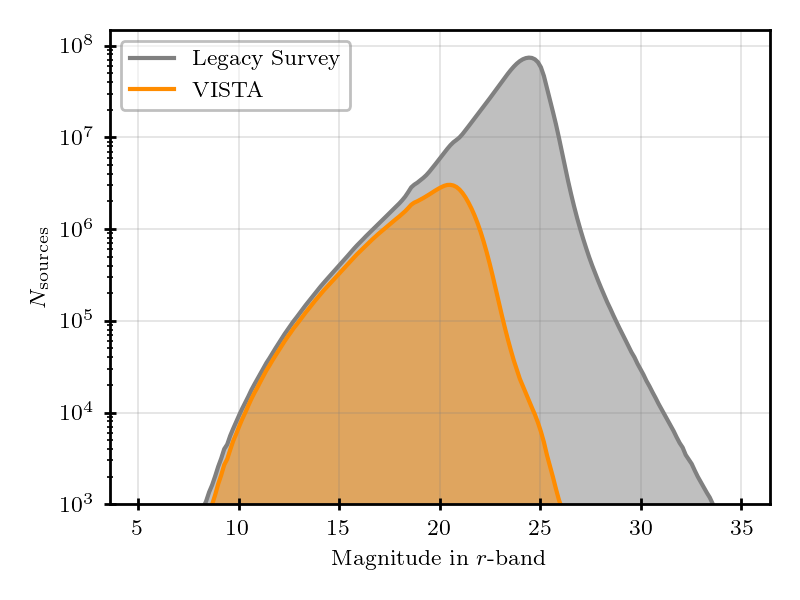}
    \vspace{-0.8cm}
    \caption{Histogram of the $r$-band magnitude from the DESI Legacy Survey (grey) with the corresponding matched VISTA sources (orange).}
    \label{fig:sources_LS_VISTA}
\end{figure}




\begin{table*}
    \centering
    \caption{Previous 4MOST-CRS target selection using VHS, VIKING and DES for BG, LRG, and ELG.}
    \begin{tabular}{c|c|c|c c}
        \hline
        \hline
        Sample  & Photometric & Mag Range & \multicolumn{2}{c}{Colour Selections}\\
        & Surveys & (AB) & & \\
        \hline
        \multirow{2}{*}{BG} & VHS+VIKING & \multirow{2}{*}{$16<J<18.25$} & $J - W1 > 1.6 \, \times \, (J - Ks) - 1.6,$ & $J - W1 < 1.6 \, \times \, (J - Ks) - 0.5$ \\
        & WISE & & $J - W1 > -2.5 \, \times \, (J - Ks) + 0.1$ & $J - W1 < 0.5 \, \times \, (J - Ks) + 0.1$ \\
        \hline
        \multirow{2}{*}{LRG} & VHS+VIKING & \multirow{2}{*}{$18 < J < 19.5$} & $J - W1 > 0.5 \, \times \, (J - Ks) + 0.05$ & $J - W1 < 0.5 \, \times \, (J - Ks) + 0.5$ \\
        & WISE & & \multicolumn{2}{c}{$J - W1 > -0.5 \, \times \, (J - Ks) + 0.1$} \\
        \hline
        \multirow{2}{*}{ELG} & \multirow{2}{*}{DES} & \multirow{2}{*}{$22.0 < g < 23.2$} & $0.5 - 2.5 \, \times \, (g - r) < (r -i)$ & $3.5 - 2.5\,\times\,(g - r) > (r - i)$ \\
        & & & $(r - i) > 0.4 \, \times \, (g - r) + 0.3$ & $(r - i) < 0.4 \, \times \, (g - r) + 0.9$ \\
        \hline
    \end{tabular}

    \label{tab:VISTA_cuts}
\end{table*}

\subsection{Change from VISTA to Legacy Survey}
\label{sec:vista_to_ls}

The photometric survey used for target selection has been switched from VISTA to DESI Legacy Surveys DR10.1 for the following reasons:

\begin{enumerate}
    \item \textbf{Leveraging cross-correlations with DESI}:
    Using DESI Legacy Survey photometric data with a similar target selection as in DESI will allow a combination of DESI and 4MOST-CRS datasets, providing a unique opportunity to construct a large spectroscopic dataset over $\sim21~000$~deg$^2$ for cosmological analysis. As shown in \cref{sec:forecasts}, even if the density of the targets is smaller, measurements of the BAO scale and the growth rate of structure $f\sigma_8$ are improved when considering a combined dataset. In addition, the quality of the target selections presented in this paper can be cross-checked with publicly available spectroscopic data from the DESI DR1 (see \cref{sec:DESI_DR1}) especially regarding the star--galaxy separation. Finally, recent results from DESI showed some deviations from previous spectroscopic surveys in the LRGs sample~\citep{DESI_BAO_DR1, DESI_BAO_DR2}. Having a similar selection in the southern hemisphere but with a higher quality of photometric data from DES will provide a strong point of comparison with DESI data.

    \item \textbf{Quality of the photometry}: The DESI Legacy Survey (LS) has been designed for spectroscopic follow-up operations. Therefore, it provides all the information needed on observation specificities to ensure the quality of cosmological analysis, which was not the case for VISTA. As an example, LS provides all the necessary information to model the systematic dependencies of the target selection with respect to photometric features such as the galaxy depth.
    We matched VISTA and LS targets via their coordinates (requiring a separation $<1''$) to compare the depth of both samples in each band. \cref{fig:sources_LS_VISTA} shows $r$-band magnitudes of the sources measured by LS (gray) with the matched VISTA objects (orange). The sources from LS display clearly deeper magnitudes compared to the matched VISTA sources. In \cref{tab:phot_info}, the $Y$ band from VISTA is deeper than for LS, which is expected as VISTA is a near-infrared survey, but we get a less significant difference than for the $r$-band. The $z$-band as a similar depth between the 2 surveys. Overall, we find that LS allows observation of fainter sources with better quality, mostly due to the DES depth.
    Another inconvenience from the VISTA photometry is that the difference in depth between observation tiles contributed to inhomogeneity in the sky selection depth of the sources. In addition, the footprint has small non-observed regions, which makes the survey window hard to model. In contrast, the LS provides full information on survey footprint and completeness as shown in \cref{fig:LS_footprint}. Moreover, LS has the advantage of providing a set of random catalogues that models the survey window and selection function for clustering analysis. 
\end{enumerate}

Regarding the survey footprint, one major update has been made because of the switch to LS. In the South Galactic Cap (SGC), the region from declination -15$^\circ > \delta > -39^\circ$ with right ascension $\alpha > 322^\circ$ has been removed as it was not completely observed in the Legacy Survey DR10.1. To compensate, the North Galactic cap (NGC) footprint has been extended in the south to match the Legacy Survey footprint instead of keeping the ATLAS footprint.
Similarly, the SGC footprint has also been extended following the Legacy Survey footprint. In the eastern part of the NGC region of 4MOST-CRS, we add an additional cut at $\alpha < 235^\circ$ as this region suffers from high galactic extinction.

\begin{figure*}
    \label{fig:LS footprint}
    \centering
    \includegraphics[width=\textwidth]{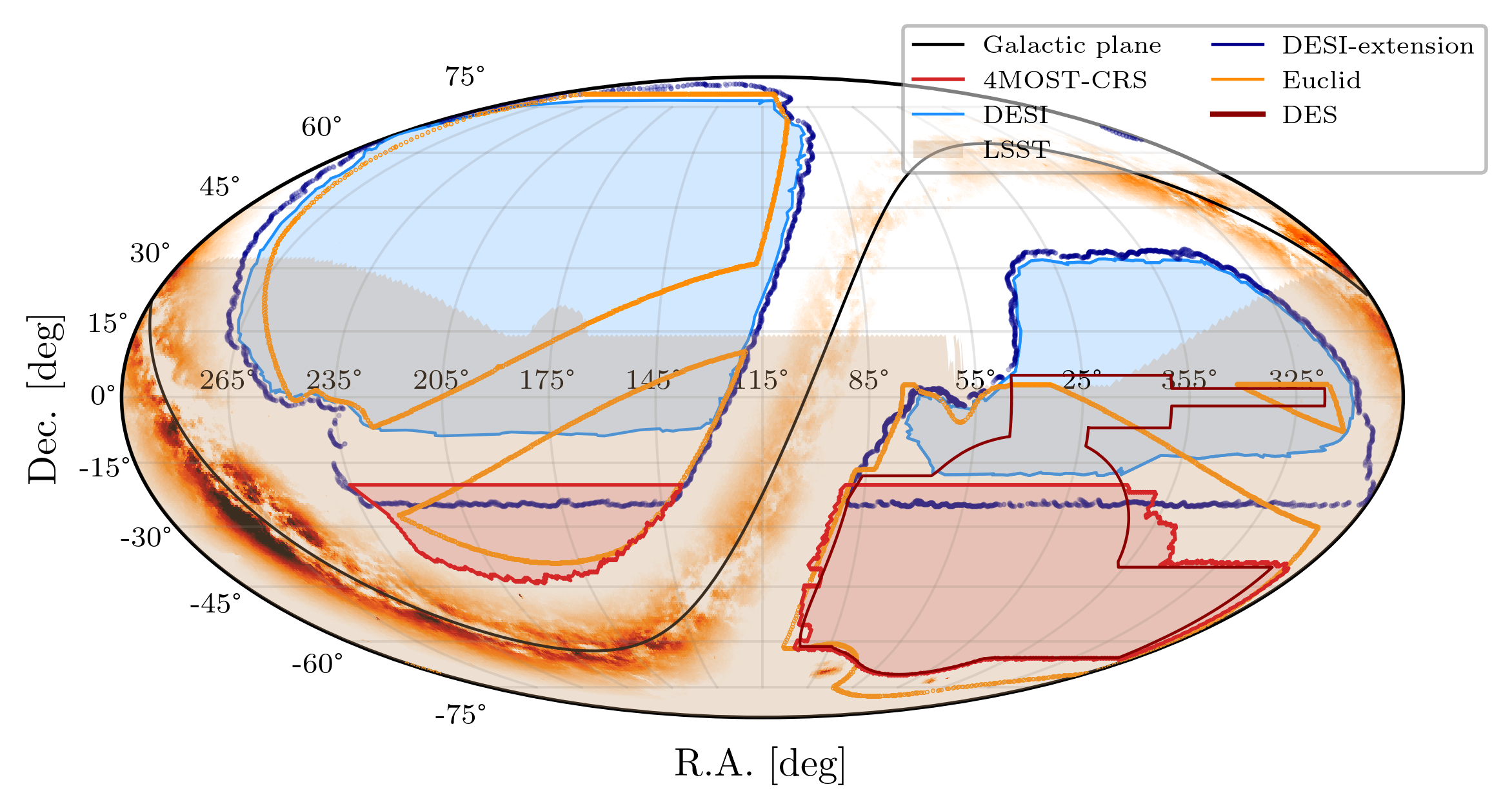}
    \vspace{-0.5cm}
    \caption{Sky map displaying the star density across the sky and different relevant surveys for 4MOST-CRS. The 5700 {\degsq} footprint of the 4MOST-CRS is represented in red. The dark blue contours represent the final footprint of the pre-extended DESI survey, while the DESI extension is represented in light blue. \textit{Euclid} and LSST footprints are also displayed, respectively, in green and grey. Overlapping sky coverage with \textit{Euclid} in the SGC is 4216{\degsq} and 1020{\degsq} in the NGC.}
    \label{fig:footprint}
\end{figure*}

Finally, a cut requiring declination $\delta < -20^\circ$ in the NGC is applied to avoid overlap with the planned DESI-extension to this declination.
The area covered by 4MOST-CRS BGs and LRGs samples is around 5700\,deg$^2$.
Compared to the former footprint built with VISTA, our LS-selected target sample loses an area of $\sim$500\,deg$^2$ on the sky\footnote{These solid angle estimates are conservative in that they have been computed without taking into account the non-observed region of the VISTA footprint, so the actual loss is less than 500\,deg$^2$.}, but is likely to yield more robust results in cosmological analysis (see \citet{Bandi_in_prep} for more details).
The new CRS footprint is shown in \cref{fig:footprint} alongside the footprints of DESI, DESI-extension, LSST and \textit{Euclid}.
As a side note, Legacy Survey DR11 should cover the missing area in the SGC region. However, the expected release date will follow 4MOST commissioning and thus also take place later than the target catalogue finalisation by 4MOST surveys.
To compare the targets selected with the new selection and the VISTA one, we match the targets between the VISTA and Legacy Survey catalogues
Targets are flagged as common if their sky coordinates agree within 1 arcsecond.
A colour-colour comparison is performed on the matched targets and shown in \cref{fig:lrg_vista}.
For BGs, the sources selected with the new LS selection (in red) lie in the same but broader region of the colour--colour plot compared to the VISTA selection (in orange). For LRGs, the new selection deviates from the original selection, but the targets still lie in the same area of colour space.




\section{Target selection from DESI Legacy Surveys}
\label{sec:LS_target_sel}

\subsection{Legacy Survey DR10.1} \label{sec:LS_phot}
The Legacy Survey DR10.1 is made of three different surveys, two in the northern celestial hemisphere (MzLS,\citealt{mzls2016} and BASS, \citealt{Zou2017BASS}) and one in the southern hemisphere (Dark Energy Camera Legacy Survey, DECaLS, \citealt{decals2015}).
All surveys use the $g$ (470\,nm), $r$ (623\,nm) and $z$ (913\,nm) optical bands.
Moreover, the survey is combined with the mid-infrared imaging WISE ($W1$, $W2$, $W3$ and $W4$, respectively centred at 3.4, 4.6, 12 and 22\,$\mu$m) and the Dark Energy Survey (DES).
As DES used the same camera as DECaLS (Dark Energy Camera, DECam), the Legacy Survey processed the public raw data with their own pipeline to include them in their catalogues.
The difference between DECaLS and DES is the depth, as the latter goes significantly deeper in each band (\cref{tab:phot_info}), which allows better photometry quality. This is confirmed in \citealt{Bandi_in_prep} where they compared the angular clustering in the north and south fields (corresponding respectively to DECaLS and DES).
4MOST will observe in the southern hemisphere, which implies that the targets will be selected only in the DECaLS survey, combined with DES and WISE.
DECam is located at the Cerro Tololo Inter-American Observatory on the Blanco 4m telescope and has observed a total area of 9000\,deg$^2$.
In total, DECaLS and DES cover $~\sim$14~500\,deg$^{2}$ over the full sky.
\begin{figure}
    \centering
    \includegraphics[width=\linewidth]{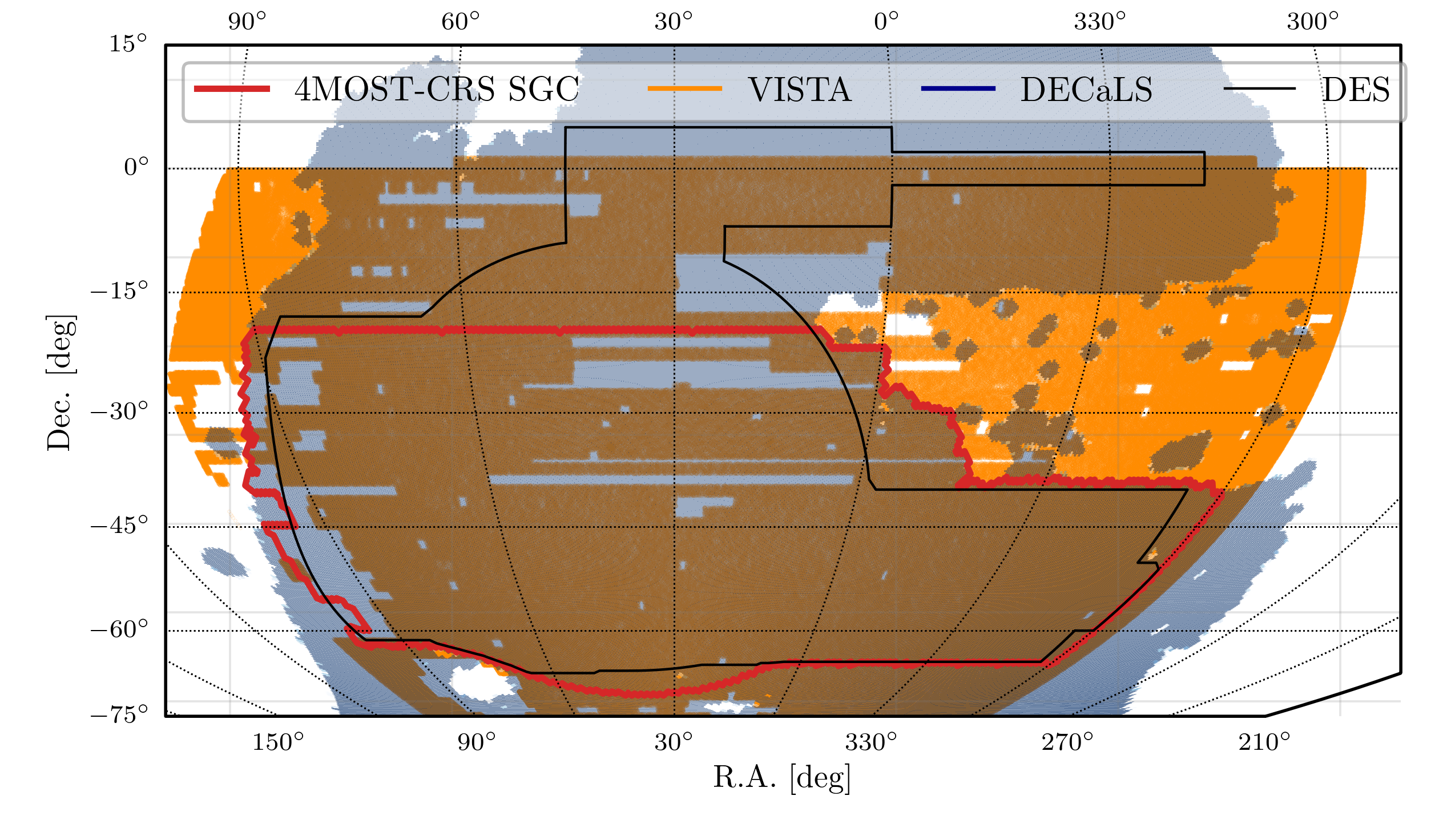}
    \vspace{-0.5cm}
    \caption{Skyprojection in a part of the South Galactic Cap. The red and black lines correspond to the 4MOST-CRS and DES footprints, respectively. The VISTA and DECaLS footprints are shown in orange and blue, respectively. The VISTA footprint clearly shows a large number of non-observed regions.}
    \label{fig:LS_footprint}
\end{figure}

\subsection{4MOST-CRS Target Selection: BGs and LRGs} \label{sec:LS_TS}

The target selection of the BGs and LRGs is inspired by the approach of the DESI survey~\citep{Zhou_2023, Hahn2023}. We used the photometric information of the $grz$ bands from the public data release DR10.1 of the Legacy survey~\citep{legacy_survey} and the $W1$-band from the Wide-field Infrared Survey Explorer (WISE) survey~\citep{WISE2010}. The full Legacy Survey DR10.1, with at least one photometric observation in each of the $grz$ bands, covers $\sim 20~000$ deg$^{2}$ of the sky. The footprint of 4MOST-CRS is defined from the DR10.1 LS in the southern part of the sky, represented in red in \cref{fig:LS_footprint}. The footprint removes declinations north of $\delta = -20^\circ$ to reduce the overlap with the DESI-ext survey. The 4MOST-CRS sky area is around $\sim 5700$ deg$^{2}$.
The next two sections precisely describe both the BGs and LRGs target selections, and a summary of both sample selections can be found in \cref{tab:cut_table}.
Given the number of fibre hours allocated to the 4MOST-CRS, additional apparent magnitude and colour cuts have been made to reduce the number of targets per square degree, reaching $\sim 250$ BGs and $\sim 400$ LRGs per square degrees compare to the DESI target selection reaching 864\,deg$^{-2}$ and 605\,deg$^{-2}$ for BGs and LRGS respectively.

\subsubsection{Bright Galaxies}
\label{sec:BG_selection}
The BG target selection used the photometric data from the LS DR10.1, the Gaia Early Data Release 3 catalogue and the Tycho-2 Bright Star catalogue. The first part of the CRS-BG target selection is similar to the one from the DESI bright galaxy sample (BGS) described in~\citep{Hahn2023}. We then apply additional cuts to remove stars based on GAIA proper motions and parallaxes, add $g-r$ vs $r-z$ colour cuts to remove low redshift targets (based on photometric redshifts) and reduce the magnitude limit to $r < 19.25$ to match the required target density. The details of this target selection process are described below.

We first apply spatial masking for objects around bright stars with magnitude \texttt{MAG\_VT}~$< 13$ for Tycho-2 sources and $G < 13$ for Gaia stars using the \texttt{BITMASK} 1 from the LS catalog\footnote{\url{https://www.legacysurvey.org/dr10/bitmasks/}}.
Moreover, depending on the magnitude of the star, we also take out more pixels using \texttt{BITMASK} 11.
This mask was not used in DESI BGS target selection but was added after seeing a reduction of the sample contamination the angular clustering analysis~\cite{Bandi_in_prep}.
Finally, we apply a spatial masking around large galaxies from the Siena Galaxy Atlas 2020 (SGA-2020) catalogue~\citep{Siena2020} (\texttt{BITMASK} 12) and around globular clusters (\texttt{BITMASK} 13).

For the sources in the Gaia EDR3 catalogue, a criterion is applied to perform the star--galaxy separation:
\begin{equation}
    G_{\textrm{Gaia}} - r_\textrm{raw} > 0.6,
\end{equation}
where $G_{\textrm{Gaia}}$ is the G-band magnitude from the Gaia EDR3 catalogue and $r_\textrm{raw}$ is the LS $r$-band magnitude that is not corrected for galactic extinction.
To remove potential spurious targets or fragments of ‘shredded’ galaxies, we apply the fibre magnitude cut (FMC) below:
\begin{equation}
r_{\text {fibre }}< \begin{cases}22.9+(r-17.8) & \text { for } r<17.8 \\ 22.9 & \text { for } 17.8<r<20 ,\end{cases}
\end{equation}
where $r_{\text {fibre }}$ is the predicted $r$-band magnitude of an object within a fibre of diameter 1.5'' in 1'' Gaussian seeing derived from the total $r$-band flux.

Then, quality cuts are performed. We only select regions where we have at least one photometric observation in each of the bands $grz$:
\begin{equation}
\operatorname{nobs}_i>0 \quad \text { for } i=g, r, z \text {, }
\end{equation}
and exclude spurious objects by removing sources with extreme colours:

\begin{equation}
\begin{aligned}
& -1<(g-r)<4 \\
& -1<(r-z)<4.
\end{aligned}
\end{equation}

Lastly, faint objects can be contaminated by very bright objects on neighbouring fibres. Thus, all sources that meet the following conditions are removed:

\begin{equation}
\left(r>12 \right) \& \left(r_{\text {fibretot }}<15\right),
\end{equation}
where $r_{\text {fibretot }}$ is the predicted $r$-band magnitude within a fibre of diameter 1.5'' from all sources at this location assuming 1'' Gaussian seeing.

The steps described above are similar to the DESI BGS-BRIGHT selection from~\citep{Hahn2023}. In addition to the DESI selection cuts, we add a selection criterion to remove stars in the galaxy sample and reduce the number of targets in the sky. All objects from the Gaia EDR3 catalogue with a non-null proper motion or parallax having a signal-to-noise ratio (SNR) $>3$ are removed.

We finally end up with a target density close to the DESI selection $\sim 820$ target/deg$^2$. In order to reach the required target density for the 4MOST-CRS, we apply the three additional colour selections:

\begin{equation}
\begin{aligned}
& r -z < (0.93\,\times\,(g-r)-0.27) \\
& r -z > (0.4\,\times\,(g-r) + 0.07) \\
& g - r > 0.95.
\end{aligned}
\label{eq:BG colour cut}
\end{equation}

\begin{figure}
    \centering
    \includegraphics[width=\linewidth]{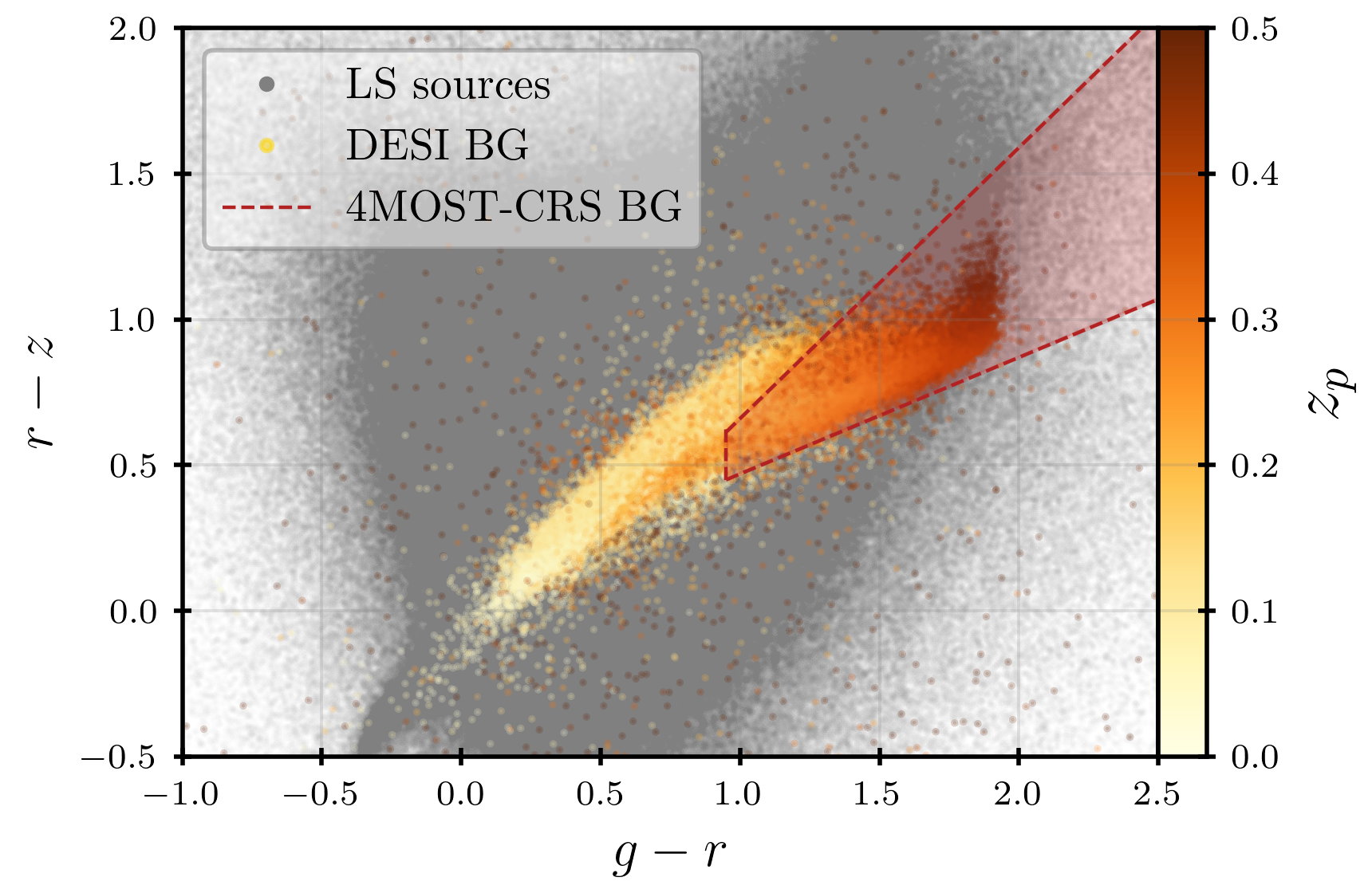}
    \vspace{-0.5cm}
        \caption{Colour-colour distribution of BGs from LS sources in the $r-z$ versus $g-r$ plane. The colour corresponds to the photometric redshift of each source derived from~\protect\cite{Zhou_2021}. The red shaded region represents the selected 4MOST-CRS BGs via the three linear colour cuts represented by the red dashed line.}
    \label{fig:BG colour selection}
\end{figure}

This colour selection has been constructed to remove objects with photometric redshifts smaller than 0.1 that are of limited use for BAO or RSD cosmological analysis and reduce the stellar contamination based on the DESI DR1 spectroscopic sample (see \cref{fig:true_gal}). \cref{fig:BG colour selection} presents the $r-z$ versus $g-r$ distribution of the LS sources, coloured by the photometric redshift derived from~\citep{Zhou_2021}. The region in which the targets are selected, defined by the selection criteria in \cref{eq:BG colour cut}, is shown shaded in red, surrounded by red dashed lines. Objects outside of this region are removed from the target catalogue, being mostly low-redshift objects ($z< 0.15$).

Finally, to reach the expected target density of $\sim 250$ targets/deg${^2}$, we impose a magnitude limit of $r < 19.25$.

\subsubsection{Luminous Red Galaxies}

\label{sec:LRG_section}

Similarly to the BGs target selection, we based our target selection on that used for the DESI LRGs as described in~\citep{Zhou_2023}. The DESI target selection of LRGs is designed to obtain a complete sample of galaxies across intermediate redshifts from 0.4 to 1.0, reaching $\sim$600 targets/deg${^2}$.
We perform some modifications to reduce the target density in order to meet the 4MOST-CRS requirements.
Our primary selection criteria are

\begin{align}
    &(z - W1) > 0.8 \, \times \, (r - z) - 0.6 \label{lrgcut_stars}\\
    &(g - W1) > 2.9 \quad \text{or} \quad (r - W1) > 1.8 \label{lrgcut_lowz}\\
    &(r - W1) > 1.8 \, \times \, (W1 - 17.08) \quad \text{and} \quad (r - W1) > W1 - 16.24 \label{lrgcut_dens} \\
    &17.5 < z_\textrm{fibre} < 21.6 \label{lrgcut_zcut},
\end{align}
where $g$, $r$, $z$ and $W1$ are the magnitudes in the AB system and are corrected for Galactic extinction using the SFD98 map (\citealt{Schlegel1998}), $z_{\textrm{fibre}}$ is derived from the expected flux measured in a DESI fibre. We also require that $G_{\textrm{Gaia}}$ is zero, meaning we remove Gaia objects.

\begin{figure}
    \centering
    \includegraphics[width=\linewidth]{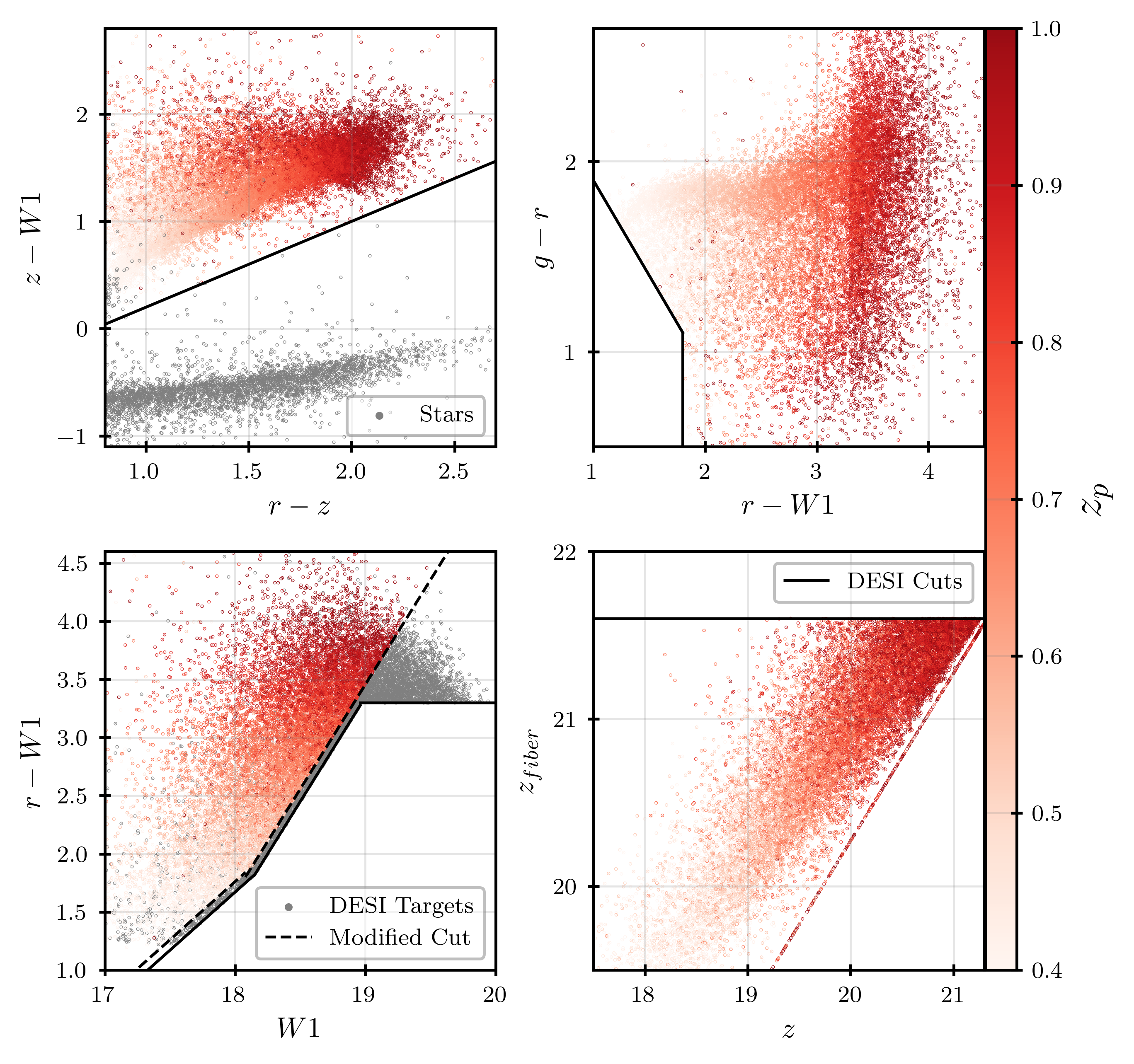}
    \caption{LRG colour-colour cuts. $g,r,z$ and $W1$ are the magnitudes derived from the flux measured in Legacy Survey DR10.1 and corrected by the galactic extinction in each corresponding band. $z_p$ is the photometric redshift estimated in the LS. In the top left panel, stars are coloured in grey to show the star--galaxy separation (stars are selected assuming $G_{\textrm{Gaia}} \neq 0$). On the bottom left, targets that were selected in DESI but are removed from our selection are coloured in grey. Residuals DESI targets (grey points) that lie in the 4MOST-CRS selection come from the additional cuts (mainly DESI selection allows target with $G_{\textrm{Gaia}}> 18$ while GAIA object are removed from the 4MOST-CRS selection). The strong black lines represent the DESI LRG cuts, while the dashed black line in the bottom left panel is the modified cut applied for 4MOST-CRS to lower the target density and the maximum redshift.}
    \label{fig:lrg_cuts}
\end{figure}

The colour cut in \cref{lrgcut_stars} removes stars from the sample using the 1.6\,$\mu$m “bump''~\citep{Sawicki_2002} in the spectrum of a galaxy within a redshift range of 0 to 1.0.
Objects classified as stars are coloured grey in the upper left panel of \cref{fig:lrg_cuts} and correspond to sources with $G_{\textrm{Gaia}} \neq 0$.
The stellar contamination of the final sample, discussed in \cref{sec:DESI_DR1}, reaches a maximum of 4$\%$ at lower magnitudes.
The colour cut in~\cref{lrgcut_lowz} limits the lower bound of the redshift range of the 4MOST-CRS LRGs sample to approximately $z > 0.4$.

While the above cuts are similar to DESI, the main difference in selection with respect to DESI LRGs target selection arises from \cref{lrgcut_dens}. This colour cut is used to select the targets as a function of their luminosity in the $W1$ band and leads to a uniform redshift distribution. Two modifications were made. DESI includes targets with $(r - W1) > 3.3$, corresponding to lower luminosity and higher redshift targets. To improve the completeness of the number of targets observed by the end of the experiment, we omit this criterion (see \cref{sec:comp}).
Secondly, we lower the target density, without modifying the redshift distribution, by shifting the overall cut to slightly brighter $W1$ and redder $r-W1$ objects compared to DESI.
The bottom-left panel in \cref{fig:lrg_cuts} shows the difference between the 4MOST-CRS LRGs and the DESI LRGs samples.
The 4MOST-CRS LRGs sample aims to have a slightly brighter faint-end limit than the DESI LRGs sample and to reach a target density of ~400\,deg$^{-2}$. The residual DESI targets (shown in grey) that fall within the 4MOST colour selection criteria but are not selected are due to additional quality criteria, primarily the exclusion of objects with $G_{\textrm{Gaia}} > 18$.
The final redshift distribution is discussed in \cref{sec:DESI_DR1}.

The cut in \cref{lrgcut_zcut} follows DESI LRGs selection, avoiding faint targets. Using the expected magnitude in a DESI fibre, our morphology selection criterion matches that of DESI.

Similarly to the BGs, we also remove objects from the Gaia EDR3 catalogue with a non-null proper motion or parallax having a signal-to-noise ratio (SNR) $>3$. Quality cuts similar to those performed for the DESI LRGs target selection are performed:
\begin{align}
&\operatorname{nobs}_i>0 \quad &&\text { for } i=g, r, z, W1 \\
&\operatorname{Flux\_ivar}_i>0 \quad &&\text { for } i=g, r, z, W1 \text {, }
\end{align}
where $\operatorname{Flux\_ivar}_i$ is the inverse variance of the flux in each photometric band.
Finally, we also perform spatial masking using \texttt{BITMASK} 1, 11, 12 and 13 as described in the BGs selection.

From the angular clustering analysis presented in~\cite{Bandi_in_prep}, all $W1$ masks from the WISE survey (\texttt{WISEMASK W1}) were added to the selection criteria. Adding this mask shows a lower contamination of the sample, yielding a two-point correlation function of the sample closer to a power law (see~\cite{Bandi_in_prep} for details). In DESI, they used a customised “veto” mask instead~\citep{DESI_2024_II}. 

\begin{table*}
    \centering
    \caption{Magnitude/colour selection for the BGs and LRGs samples. Apparent magnitudes are obtained from the variables {\tt flux\_X}, {\tt fiberflux\_X}, {\tt fibertotflux\_X}, {\tt gaia\_phot\_g\_mean\_mag}, corrected for extinction for band {\tt X}; $r_{\text{raw}}$ is uncorrected for extinction.\protect\footnote{\url{https://www.legacysurvey.org/dr10/catalogs}}}.
    \begin{tabular}{c c c r l c}
        \hline
        \hline
        Sample & Num. of & Density & \multicolumn{2}{c}{Mag. Range (AB)} & Colour Selection  \\
        & Targ. ($\times10^6$)& [deg$^{2}$] & & & \\
        \hline
          & & & \multicolumn{2}{c}{r < 19.25}  &  $G_{\text {Gaia }}-r_{\text {raw }}>0.6$ \\
          & & & & & $r-z < (0.93 \times (g-r) - 0.27)$ \\
          \multirow{2}{*}{BG} & \multirow{2}{*}{$\sim1.45$} & \multirow{2}{*}{$\sim$250} & $r_\textrm{fiber} < 22.9 + (r-17.8)$ & for $r < 17.8$ & $r-z > (0.4 \times (g-r) + 0.07)$ \\
          & & & $r_\textrm{fiber} < 22.9$ & for $17.8<r<20$ & $g-r > 0.95$ \\
          & & & & & $-1 < (g - r) < 4$\\
          & & & \multicolumn{2}{c}{$(r \le 12) \; \mathrm{or} \; (r_\textrm{fibertot} \ge 15)$} & $-1 < (r-z) < 4$\\
        \hline
          & & & \multicolumn{2}{c}{} & $z-W1 > 0.8 \times (r-z) - 0.6$\\
          LRG & $\sim2.26$ & $\sim$400 & \multicolumn{2}{c}{$17.5 < z_\textrm{fibre} < 21.6$} & $(g-W1) > 2.9 \; \text{or} \; (r-W1) > 1.8 $\\
          & & & &  & $r-W1 > 1.8 \times(W1 - 17.08) \; \text{and} \; (r-W1) > W1 - 16.24$ \\
          \hline
    \end{tabular}

    \label{tab:cut_table}
\end{table*}

\subsection{Target Selection Validation and Forecasts}
\label{sec:TS_Val}
\subsubsection{Comparison with DESI DR1}

\label{sec:DESI_DR1}
In April 2025, DESI published their first data release~\citep{DESI_DR1}, with 6’280’198 BGs and 2’829’611 LRGs, corresponding to 41.3\% and 29\% of the final DESI Main Survey for bright time and dark time, respectively~\citep{DESI_DR1}.

\begin{figure}
    \centering
    \includegraphics[width=\linewidth]{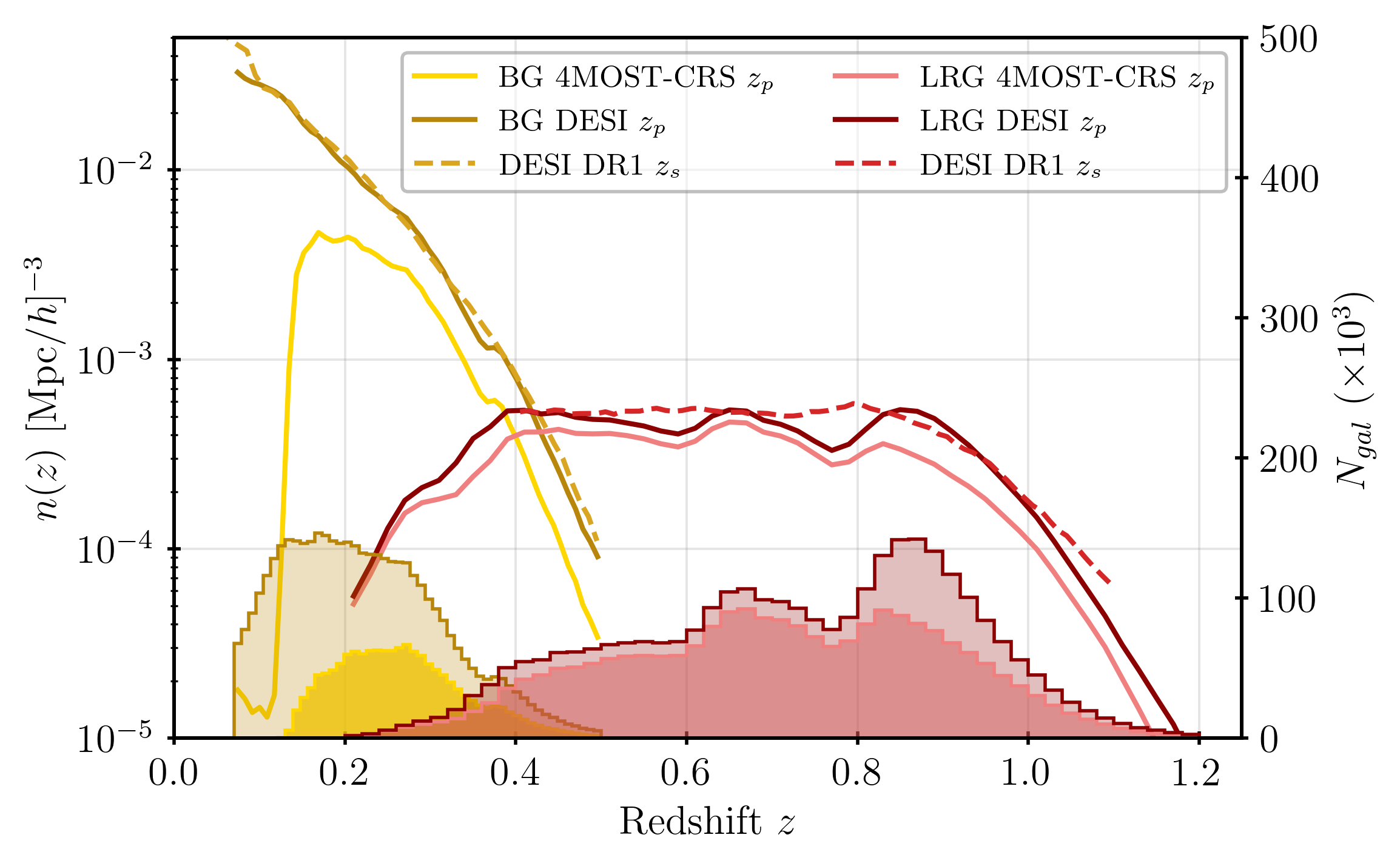}
    \vspace{-0.5cm}
    \caption{Number density (left y-axis) and number (right y-axis) of targets as a function of redshift. The BGs sample is represented with the gold and dark gold curves and the LRGs sample with the coral and dark red for the 4MOST-CRS and DESI selected targets, respectively. Lines are related to the Number density (left axis), and histograms to the total number of objects (right axis). Darker colours correspond to DESI targets. Comparatively, the number density of DESI DR1 objects are represented by the dashed line with the same colour code as DESI data.}
    \label{fig:DESI_4MOST_nz}
\end{figure}

Using these data, we check the difference in the redshift distributions using the DESI and 4MOST-CRS selection criteria (\cref{fig:DESI_4MOST_nz}).
As expected, the shapes of the distributions between the 2 selections are similar, except for the low redshift BGs with $z \lesssim 0.2$ and the high redshift LRGs with $z \gtrsim 0.8$ due to the additional cuts in the 4MOST-CRS selection (\cref{eq:BG colour cut} and \cref{lrgcut_dens} respectively).

\begin{figure}
    \centering
    \includegraphics[width=\linewidth]{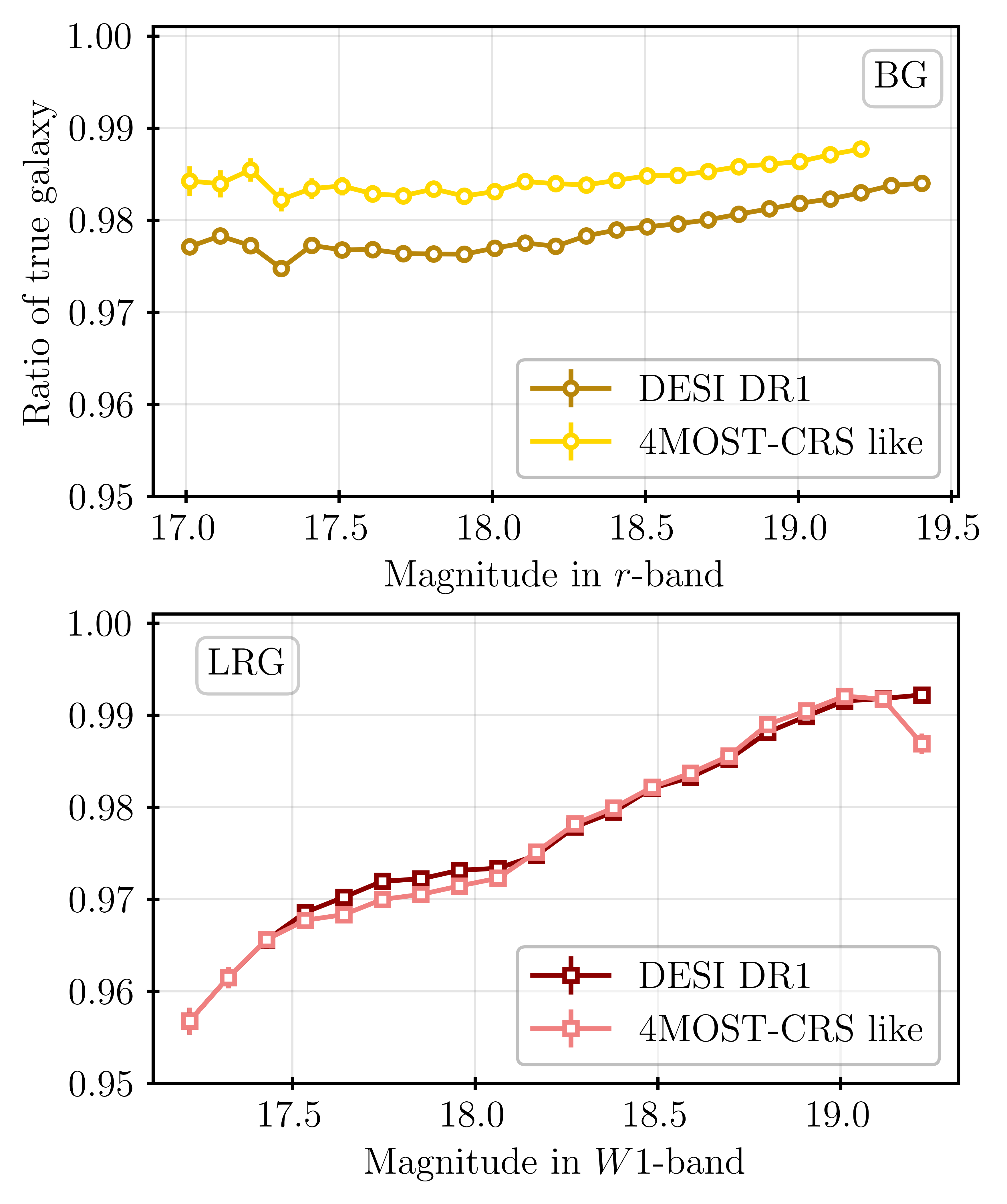}
    \vspace{-0.5cm}
    \caption{Top panel: Galaxy purity of DESI DR1 in dark gold and corresponding targets with 4MOST-CRS selection in gold for BGs. Bottom panel: Similar for LRGs, with DESI DR1 and 4MOST-CRS targets in dark red and coral, respectively. Magnitude bins are in the $r$-band for BGs and in the $W1$-band for LRGs.}
    \label{fig:true_gal}
\end{figure}

As stated above, we can check the quality of the 4MOST-CRS target selection with spectroscopically available DESI data from DR1.
By selecting the DESI DR1 targets that correspond to the 4MOST-CRS selection, we compute the ratio of spectroscopically confirmed galaxies to the targeted ones using the flag \texttt{SPECTYPE}=='GALAXY’ from the DESI DR1 catalogues.
This ratio for the two samples is shown in \cref{fig:true_gal} against $r$ for BGs and against $W1$ for LRGs.
High galaxy purity similar to DESI is achieved for both samples with a minimum of $\sim$98.2$\%$ and 96$\%$ of true galaxies among our targets, respectively, for BGs and LRGs.
The purity of BGs increases by almost 1\% in the range of $r$-band magnitude, most likely due to the addition of spatial masking and the additional colour selection compared to DESI.

\subsubsection{Samples quality check with angular clustering}
\label{sec:AC_theorie}
One way to check the quality of a sample of galaxies selected using photometric survey is to use the angular correlation function.
The two-point correlation function $\xi(r)$ is defined as a measure of the excess probability $\mathrm{d}P$ of finding a galaxy in a volume element $\mathrm{d}V$ at a separation $r$ from a randomly selected galaxy\footnote{The separation must be with respect to a randomly selected galaxy, not a randomly selected position.}:
\begin{equation}
    \mathrm{d}P = n[1 + \xi(r)]\mathrm{d}V,
\end{equation}
where $n$ is the mean number density of the galaxy sample~\citep{Peebles1980}.
Because we currently only have access to photometric redshifts, we do not have precise three-dimensional comoving-space positions of the galaxies.
Thus, we instead consider the two-dimensional case, which is the angular correlation function on the celestial sphere, replacing volumes by solid angles and distances by angular distances (Limber’s approximation,~\cite{Limber_1953}):
\begin{equation}
  \mathrm{d}P(\theta) = N[1 + \omega(\theta)]\mathrm{d}\Omega,
\end{equation}
where $N$ is the mean surface density.

\begin{figure}
    \centering
    \includegraphics[width=\linewidth]{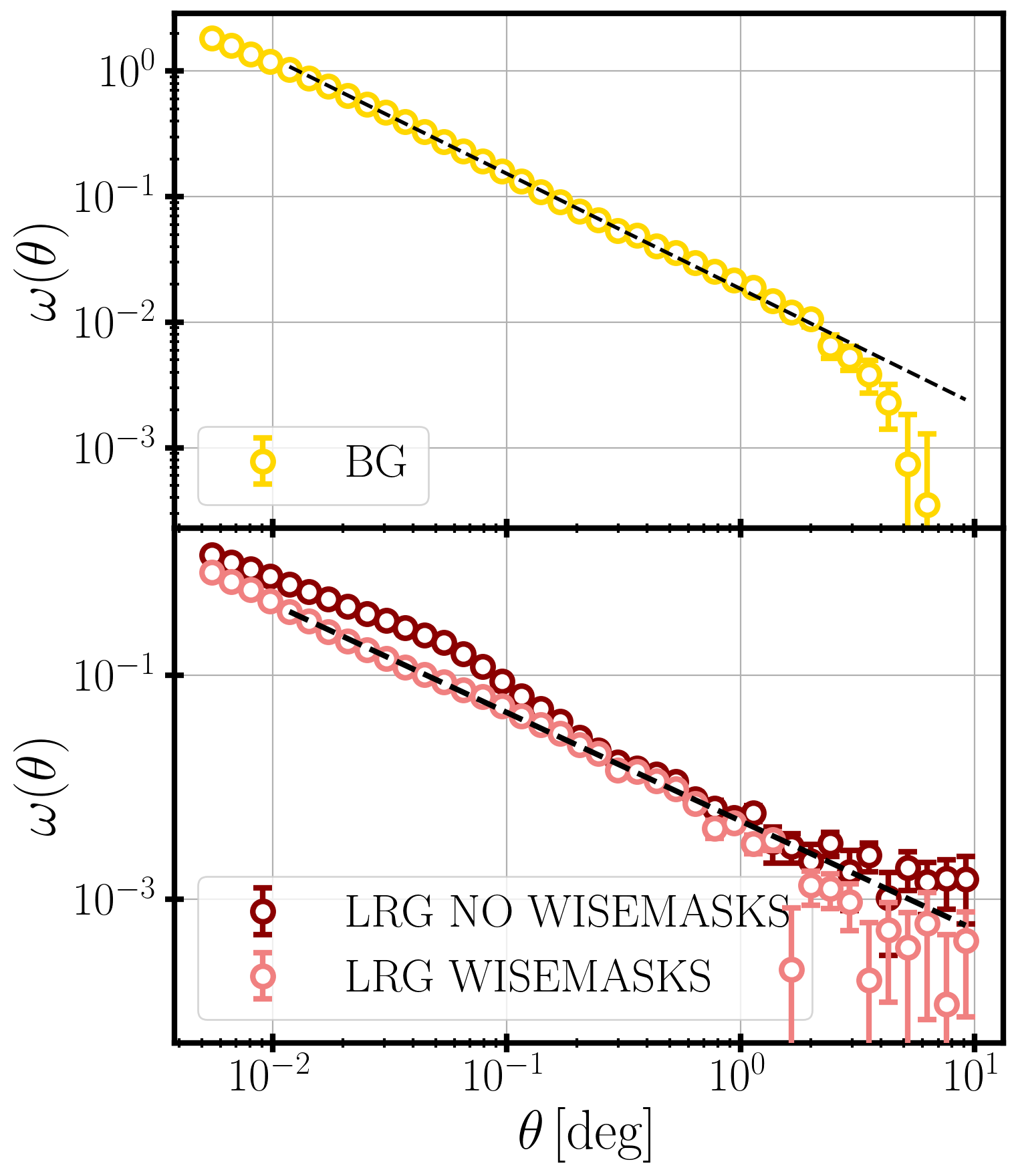}
    \caption{4MOST-CRS BGs (top panel) and LRGs (bottom panel) angular two-point functions with error bars using 36 jackknife regions. For LRGs, pink points correspond to the sample using WISE masks and dark red without WISE masks (\texttt{BITMASK} 5). A linear dashed black fit is added in both plots to highlight that $\omega(\theta)$ follows a power law.}
    \label{fig:BG_LRG_clust}
\end{figure}

We use the Landy--Szalay estimator~\citep{Landy1993} to estimate the angular correlation function $\omega(\theta)$,
\begin{equation}
    \hat{\omega}(\theta) = \frac{DD - DR + RR}{RR},
\end{equation}
where $DD$, $DR$ and $RR$ correspond to the data--data, data--random and random--random pair counts, respectively, at angular separation $\theta$.
To compute the angular correlations, we use the software package $\texttt{pycorr}$\footnote{\url{https://py2pcf.readthedocs.io/en/stable/}}, which is a python wrapper of the $\texttt{Corrfunc}$ code~\citep{Corrfunc}.
The angular two-point correlation function is often fit with a power law of the form:
\begin{equation}
    \omega(\theta) = A_{\omega}\theta^{1-\gamma},
\end{equation}
where $A_{\omega}$ is the clustering amplitude and $1-\gamma$ is the slope of the correlation function.
As pointed out by \cite{Peebles1980}, checking that the angular correlation function of galaxies follows a power law form provides a sanity check for the quality of the sample.
If a photometric target catalogue is contaminated by stars or has other systematic errors, deviations from the power law can be expected.
The resulting angular clustering functions for the 4MOST-CRS BGs and LRGs samples are shown in \cref{fig:BG_LRG_clust} in gold and pink, respectively.
We also plot the angular clustering of the LRG sample without applying WISE masks to show that without those, we do not follow a power law for $\omega(\theta)$.
We interpret this to mean that without WISE masking, the sample is polluted by stars or artefacts that hide the cosmological signal.
See \cite{Bandi_in_prep}, including more analysis of the 4MOST-CRS samples, for details.

\subsubsection{Completeness of the sample}
\label{sec:comp}
A software tool to simulate the 5-year survey of 4MOST has been developed within the consortium by the Operation System team (OpSys) and is called the 4MOST Operation Simulator (OpSim)~\citep{Tempel2020a, Tempel2020b}.
This tool helps the 4MOST survey teams check if their target catalogues are likely to achieve their science goals and provides observational information for each target.
To parametrise the success of the experiment, OpSim uses a Figure of Merit (FoM), which ranges from 0 to 1.
The FoM is defined by each survey and needs to reach a minimum score of 0.5 to be labelled as successful.
To generate the tiles in the sky that should be observed, the Visit Planner algorithm uses a probability density:
\begin{equation}
    p(\mathbf{y} \mid \theta) \propto \exp [-U(\mathbf{y} \mid \theta)],
\end{equation}
where $-U(\mathbf{y} \mid \theta)$ is the energy function which takes into account different parameters to optimise the sequence of observations.
In particular, one of the components of the energy function is the exposure time of the targets:
\begin{equation}
    U_{\text {targets }}(\mathbf{y} \mid \theta)=\frac{1}{A(\mathrm{FoV})} \iint_S U_{\text {targets }}^s(\mathbf{y} \mid \theta) \,\mathrm{d} s,
\end{equation}
where $S \in W$ is the region of the sky where targets are located and $A(\mathrm{FoV})$ is the area of one field of view.
$U_{\text {targets }}^s(\mathbf{y} \mid \theta)$ depends on the remaining exposure time needed to observe all targets, and observational time that is not used for science targets. Both parameters are weighted by a constant to fine-tune the balance between \enquote*{missing} and \enquote*{wasted} observations.
Using this model, the simulator is able to maximise the exposure time of all the different targets provided by the different surveys composing 4MOST.

\begin{figure}
    \centering
    \includegraphics[width=\linewidth]{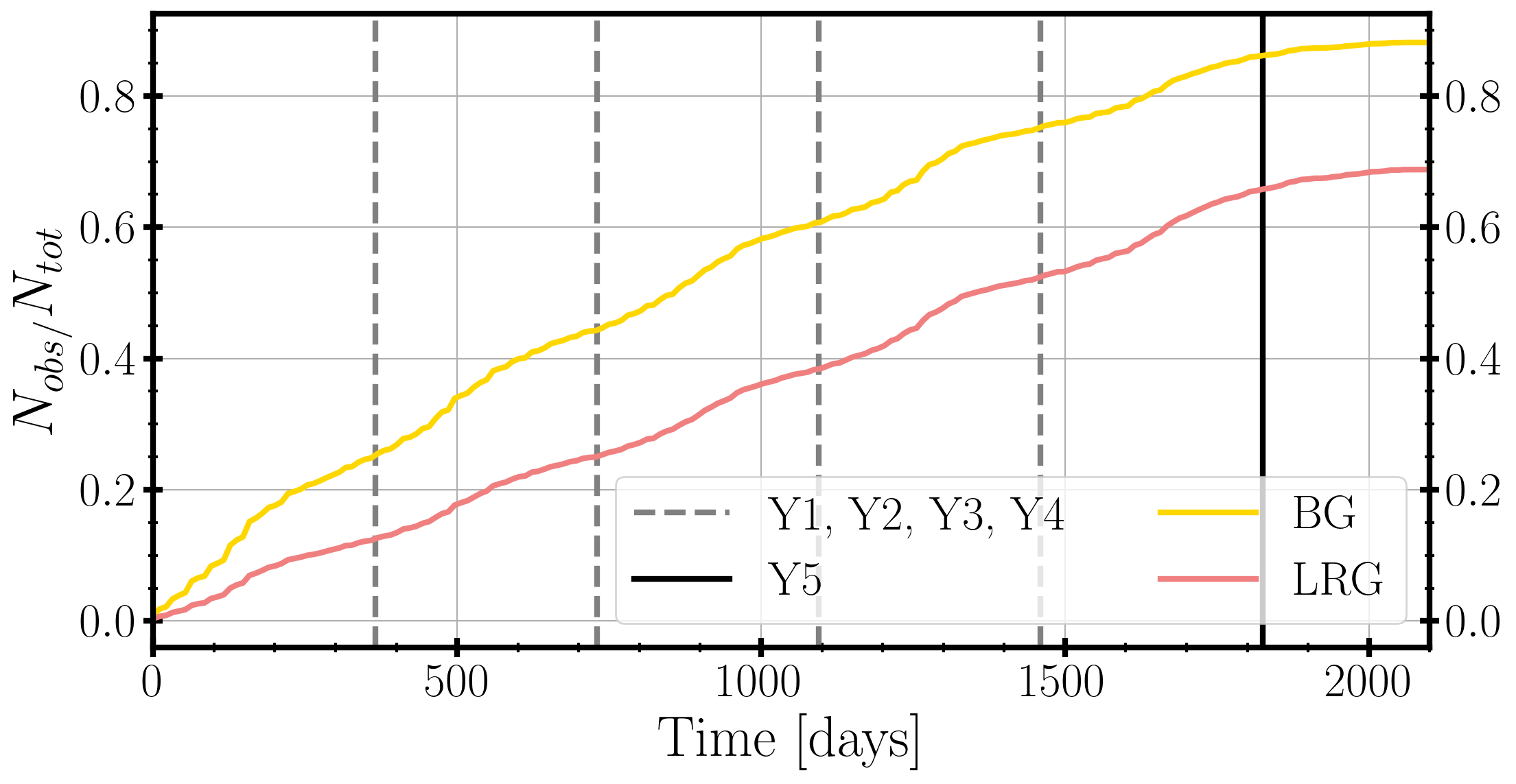}
    \caption{Average target progress as a function of survey time for BGs (yellow) and LRGs (pink). Y1-5 represents the five years for the 4MOST survey. BGs and LRGs samples respectively achieve 87$\%$ and 67$\%$ completeness at the end of the five years. The 4MOST consortium expects to have one data release per year.}
    \label{fig:comp_vs_time}
\end{figure}

Using the catalogue output of OpSim, we show in \cref{fig:comp_vs_time} the evolution in time of the number of targets successfully observed.
After 5 years of data acquisition, the completeness of the sample is expected to reach 87\,$\%$ for the BG sample and 67\,$\%$ for the LRG sample, corresponding to a FoM of 0.96 and 0.5, respectively.
4MOST-CRS satisfies the science quality criteria FoM $> 0.5$ for both the BGs and LRGs samples, even if the LRGs sample is on the threshold limit.

\subsubsection{Forecasts}
\label{sec:forecasts}
To model how 4MOST-CRS will constrain the BAO and $\sigma_{8}$ cosmological parameters, Fisher forecasts are produced using FishLSS~\citep{Sailer2021} and then compared with DESI.
The parameters needed to produce the forecasts are the fraction of the sky observed by the survey ($f_{\textrm{sky}}$), the redshift density of the targets (\cref{fig:DESI_4MOST_nz}), the number of bins for each redshift range ($n_{\textrm{bins}}$) and the bias of each sample.
To obtain the latter, we use the linear bias $b_{\textrm{lin}}$ obtained for DESI BGS in~\cite{DESI} and in~\cite{Yuan_HOD_LRG_2024} for LRGs.
For LRGs, $b_{\textrm{lin}}$ is computed in three different redshift bins: $0.4<z<0.6$, $0.6<z<0.8$ and $0.8<z<1.0$.
The values $b_{\textrm{lin}}$ for these redshift ranges are 1.93, 2.08 and 2.31, respectively.
For BGs, $b_{\textrm{lin}}$ = 1.34 is used for the whole redshift range $0.1<z<0.4$.
We assume that $b_{\textrm{lin}}$ for 4MOST-CRS galaxies is similar to that of the DESI objects. However, we might expect a higher $b_{\textrm{lin}}$ for 4MOST-CRS objects as the selection chooses brighter objects but keeps the $b_{\textrm{lin}}$ of DESI objects in order to be more conservative. 
We use the completeness obtained in \cref{sec:comp} to lower the 4MOST-CRS density of both BGs and LRGs uniformly across the sky. In practice, 4MOST observations are not uniform across the sky. This is because of the overlap between the various 4MOST surveys, which results in an inhomogeneous observed target density. It will be necessary to carefully analyse this in order to obtain unbiased clustering measurements.

DESI BGs and LRGs are observed over a total area of around $\sim$14~200\,deg$^{2}$, which corresponds to $f_{\textrm{sky}}= 0.34$.
We also take into account the DESI extension for both samples, which corresponds to $\sim$18~000\,deg$^{2}$, i.e. $f_{\textrm{sky}} = 0.44$.
With a footprint of $\sim$5700\,deg$^2$, the 4MOST-CRS BGs and LRGs cover a sky fraction of $f_{\textrm{sky}}$ = 0.14.

We also present a combined sample of 4MOST-CRS and DESI (called \enquote*{4MOST+DESI}) to show how both samples can work together to improve the constraints.
We select all targets in both the DESI and 4MOST footprints that follow the 4MOST-CRS target selection.
The value of $b_{\textrm{lin}}$ is the same as before for each redshift range, but the value of $f_{\textrm{sky}}$ increases to 0.58 in this case.

\begin{figure}
    \centering
    \includegraphics[width=\linewidth]{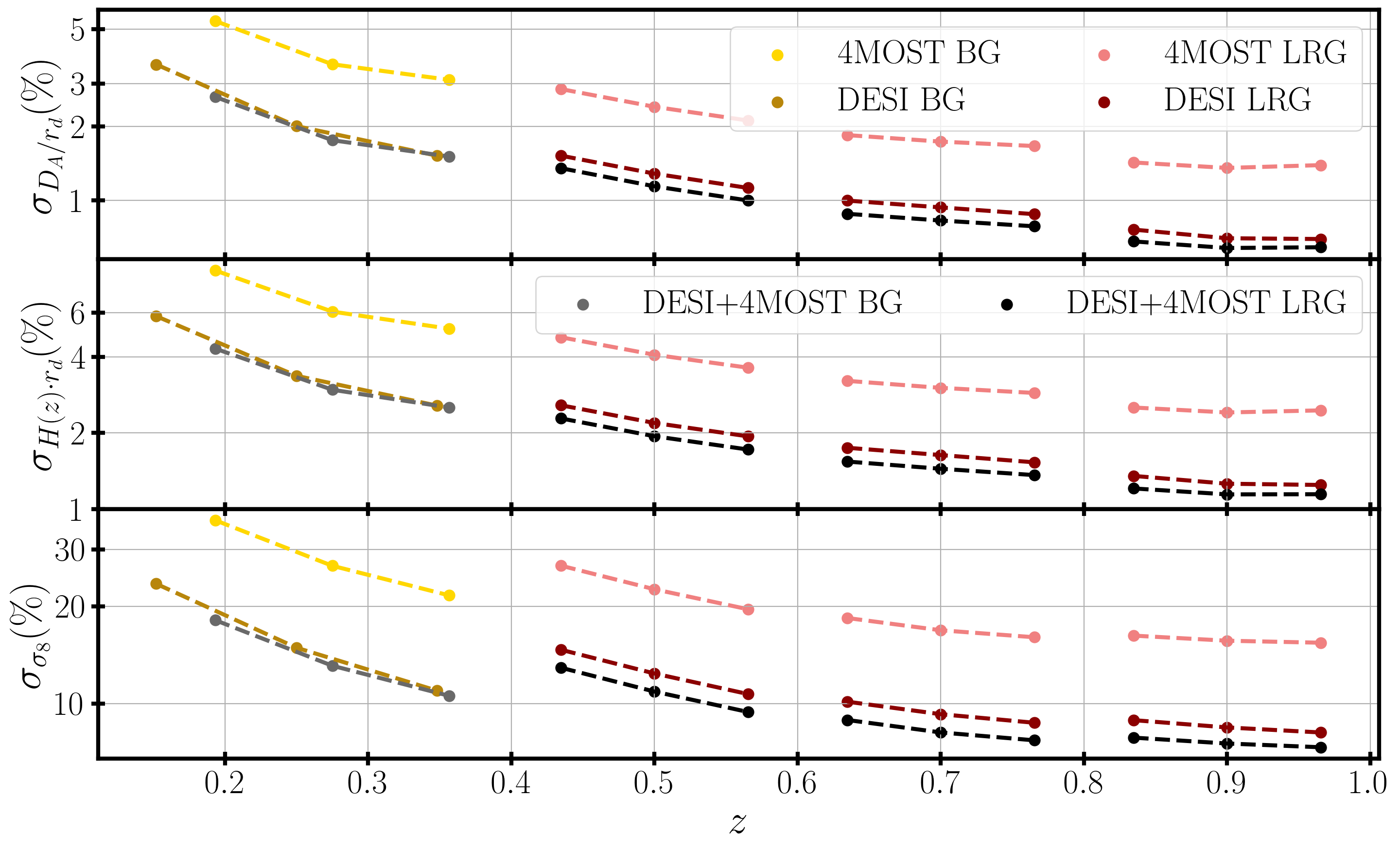}
    \caption{BAO and $\sigma_8$ Fisher forecast for the BGs and LRGs samples and bins. BGs samples are selected in the 0.1 to 0.4 redshift range in three bins. BGs from 4MOST-CRS, DESI and the combination of the two surveys (4MOST+DESI) are shown in yellow, gold and grey, respectively. The LRGs sample is split into three redshift bins: 0.4 to 0.6, 0.6 to 0.8 and 0.8 to 1.0 and evaluated at 3 redshifts for each bin. LRGs from 4MOST-CRS, DESI and 4MOST+DESI are shown in pink, purple and black, respectively.}
    \label{fig:forecasts}
\end{figure}

As expected, \cref{fig:forecasts} shows that while 4MOST-CRS is complementary to DESI in surveying a different part of the sky, 4MOST-CRS alone will not be competitive compared to DESI.
However, the combination of the two surveys will improve the constraints on cosmological parameter estimates up to $\sim12\%$ in the range $0.4 < z < 0.8$. Regarding $z < 0.4$, 4MOST+DESI will only constraint up to $\sim 1\%$, due to the low target density of the 4MOST-CRS BGs sample. At $z > 0.8$, as high redshift LRGs are dropped from the 4MOST-CRS selection (\cref{lrgcut_dens}), the improvement is reduced to $\sim9\%$.

\section{Photometric systematics} \label{sec:phot_sys}

\subsection{Systematic errors} \label{sec:sys_err}

The aim of the systematic mitigation is to correct for spurious
density fluctuations in the target selection without suppressing the
cosmological signal.

Photometric data in DECaLS and DES data do not perfectly probe the homogeneity of the Universe, as the number of galaxies observed in one pointing of the telescope will depend on many observational factors, such as the Moon’s brightness or the seeing.
Those effects lead to spurious signals in the clustering, leading to an increase of signal at large separations.
In order to mitigate these effects, we analyse the sources and likely errors induced by these effects. We consider different features linked to the observational properties
such as imaging depth, extinction (E(B-V)), and other physical properties that might altered the observations
The features we used are the same as the previous DESI study~\cite{Chaussidon2022} and are describe below:
\begin{enumerate}
    \item \textbf{Stellar density} [deg$^{-2}$]: The stellar density map is made of the point sources from Gaia DR3 (\textbf{Gaia DR3}).
    \item \textbf{E(B-V)} [mag]: Galactic extinction computed by~\cite{Schlegel1998} and corrected by~\cite{Schlafly2011}.
    \item \textbf{PSF depth} [1/nanomaggie]:\footnote{\url{https://www.sdss3.org/dr8/algorithms/magnitudes.php}} The 5$\sigma$ source detection limit in the $g$, $r$, and $z$ bands. For $W1$, the same source detection limit is applied on WISE data after conversion to the AB system.
    \item \textbf{PSF size} [arcsec]: The weighted average PSF FWHM in arcsec in the $g$, $r$, and $z$ bands.
    \item \textbf{Galaxy Depth} [mag]: Galaxy detection sensitivity (0.45 arcsec)
\end{enumerate}

\subsection{Homogeneity of the photometric surveys}
\label{sec:phot_surv}
\begin{figure*}
    \centering
    \includegraphics[width=\linewidth]{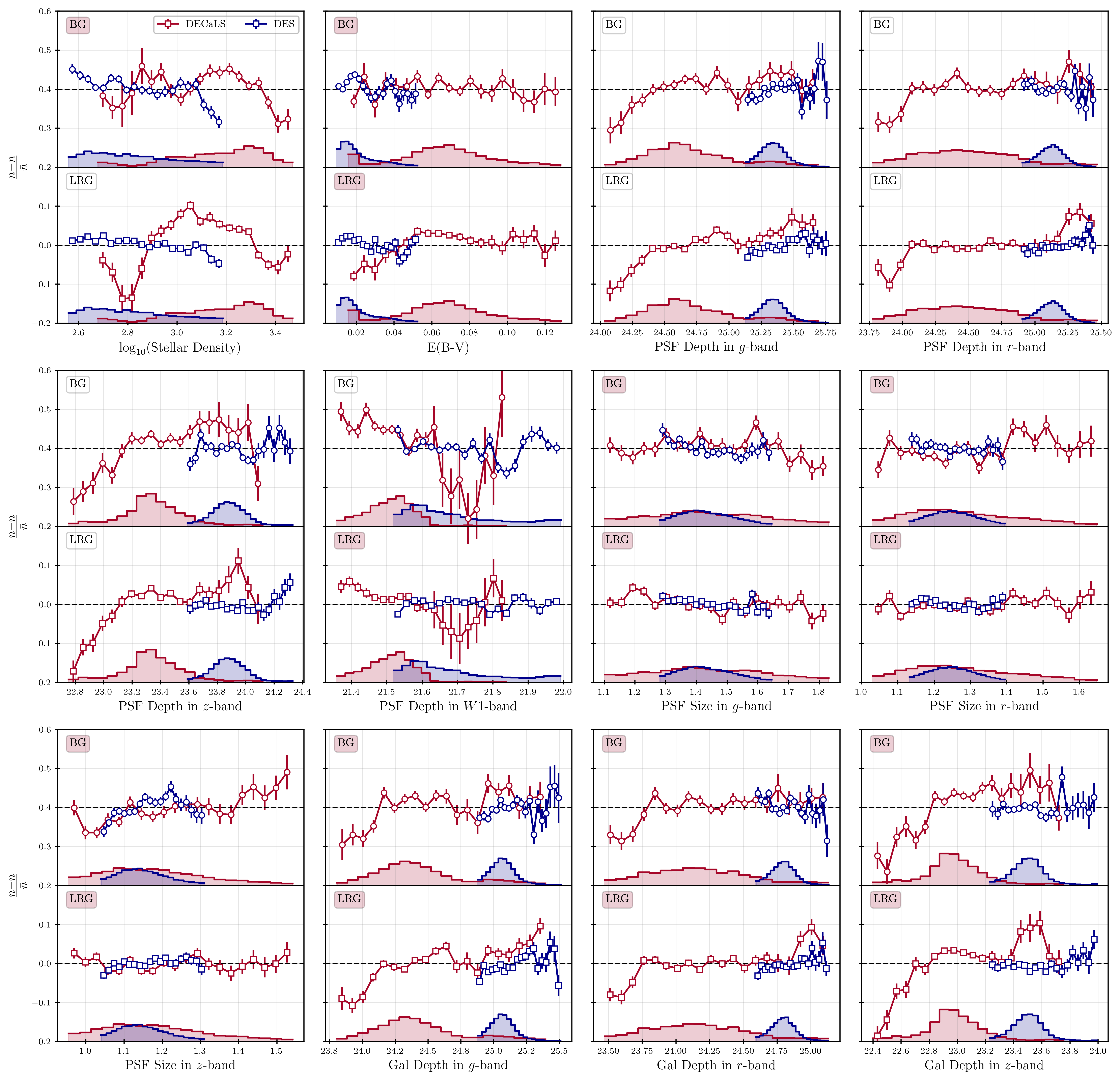}
    \caption{Relative target density as a function of each observational feature in the DES (blue) and DECaLS (red) regions. For each panel the top plot corresponds to the BG sample and the bottom to the LRG sample. The y-axis values for the BG sample has been shifted $+0.4$. The histogram represents the fraction of objects in each bin for each observational feature and the error bars are the estimated standard deviation of the normalized target density in each bin.}
    \label{fig:bg_sys}
\end{figure*}

As stated above, three photometric surveys, DECaLS, DES, and WISE, are used for target selection.
DES was not originally part of the Legacy Survey, but the public raw data have been processed by the Legacy Survey pipeline, and all parameters described above can be retrieved. As the photometric quality (depth) depends on the survey, DES being deeper than DECaLS we separate the photometric mitigation accordingly.   
For each parameter $x$, we calculate the mean number density $\overline{n_x}(x)$ of targets per square degree in bins of parameter $x$, and the overdensities with respect to these means.
The overdensities of each of these parameters in the 2 photometric regions, DECaLS and DES, are shown in \cref{fig:bg_sys}.
The main differences between the two surveys are the depth and the PSF size in the $g$, $r$ and $z$ bands and the galactic extinction $E(B-V)$.
As DES goes deeper than DECaLS with a longer exposure time, the depth and the PSF size are expected to be, respectively, deeper and narrower.
The DES footprint is farther away from the galactic centre than DECaLS, so $E(B-V)$ is less affected by Milky Way dust.
Furthermore, the two surveys have different depth distributions in the $W1$ band, as WISE made more observations near the south and north ecliptic poles.
Thus, DES shares more deep sky observations with WISE than DECaLS.
Finally, and as expected, DES is of higher quality in terms of photometry than DECaLS, giving a more homogeneous sample and a cleaner clustering signal.
Since the PSF depth and Galaxy depth are closely related, we only use the latter for systematic error mitigation similarly to~\cite{Chaussidon2022}.

\subsection{Systematic error mitigation and clustering quality check}

All the density and feature maps are pixelized using \texttt{HEALPix}\footnote{\url{https://healpix.sourceforge.io/}}~\citep{Gorski_2005} using the python package \texttt{healpy}\footnote{\url{https://healpy.readthedocs.io/en/latest/}} to produce the overdensity map of each systematic error effect described in \cref{sec:phot_sys}.
We set the size of the pixelization $N_{\textrm{side}} = 128$, which corresponds to a typical pixel size of about $\theta = 0.46$\,deg.
The number of objects per healpix is then $\sim50$ and $\sim80$ for BGs and LRGs, respectively. We based our analysis for photometric weight estimation on the \textsc{regressis}\footnote{\url{https://github.com/echaussidon/regressis}
} python package. As described in~\cite{Chaussidon2022}, the systematic correction is modelled by weight per pixel $i$ defined as:
\begin{equation}
    w_i^{\mathrm{sys}}=\frac{1}{F\left(s_i\right)},
\end{equation}
where, $F(s_i)$ is the contamination signal.
We tried two different regression methods to mitigate the photometric systematics independently for DECaLS and DES regions: linear and random forest (RF) regression.
Both regressions and their usage are described in \cite{Chaussidon2022}.
The results of the mitigation using each method for the DECaLS and DES regions are respectively shown in~\cref{app-DECaLS-DES} (\cref{fig:DECaLS_sys} and \cref{fig:DES_sys} respectively).
Regarding the DECaLS region, both methods capture and mitigate the trends in the target density according to the photometric features. As mentioned in the previous section, the photometry in the DES region is already of good quality, and uncorrected feature maps already display a small variation in the target density. Therefore, the linear or Random Forest corrections do not significantly correct the target density.

\begin{figure}
    \centering
    \includegraphics[width=\linewidth]{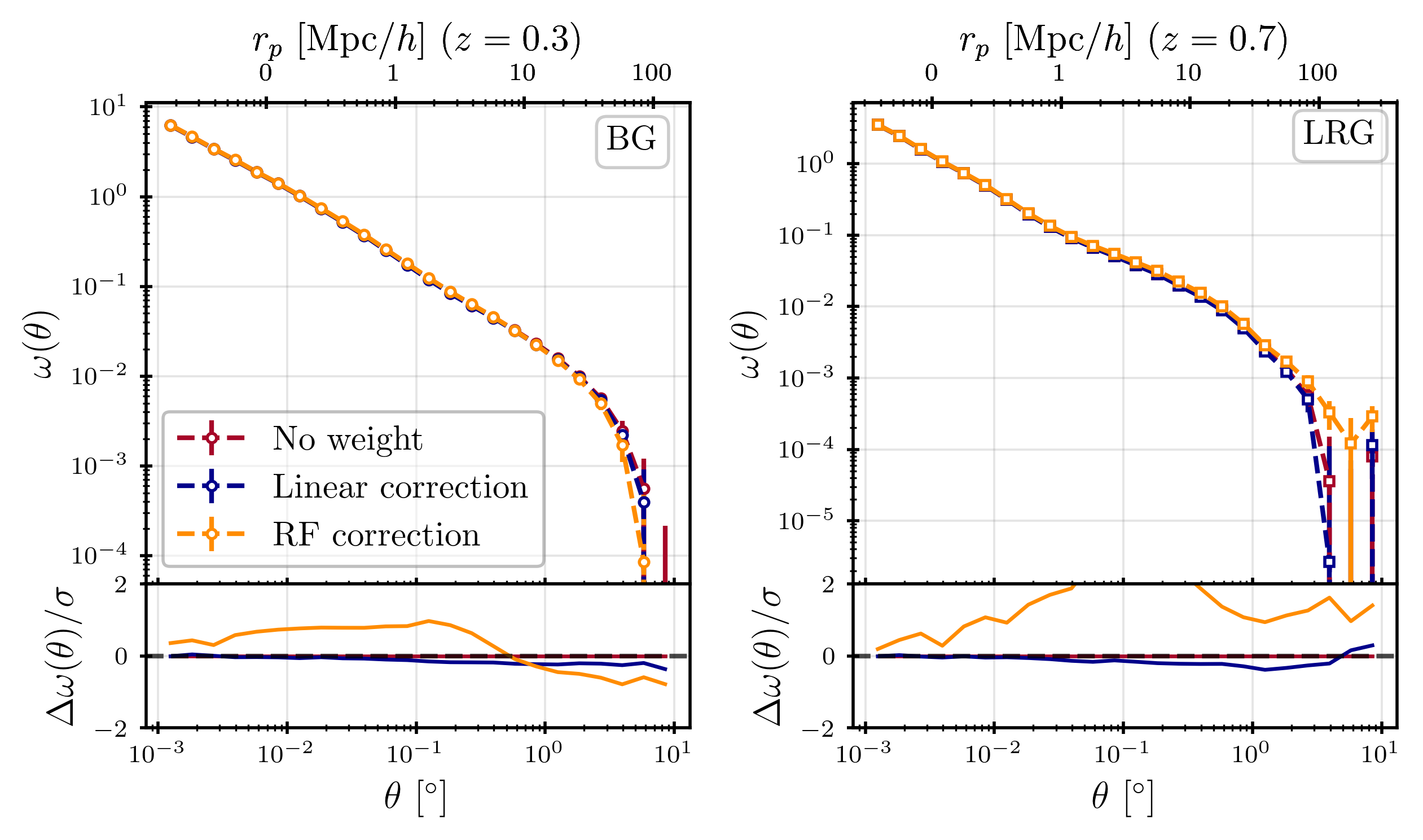}
    \caption{{\em Upper panels:} Angular two-point correlation function for BGs (left) and LRGs (right), uncorrected (no weights, red) and corrected with the linear (in blue) and the Random Forest (in orange) regression methods. {\em Lower panels:} residuals between the raw (uncorrected, red) and corrected correlation functions. The top axis display the corresponding transverse distance $r_p$ evaluate at a representative redshift for each sample ($z=0.3$ for BGs and $z=0.7$ for LRGs).}
    \label{fig:BG_LRG_w_sys_weight}
\end{figure}

Finally, to check the quality of the mitigation applied to our samples, we compute the angular clustering for both sample as described in \cref{sec:AC_theorie} using no weights, and photometric weights for both regression methods. The results are shown in \cref{fig:BG_LRG_w_sys_weight}.
For both sample, linear regression corrects better for large scale deviations while Random Forest regression shows higher corrections maybe due to overfitting.

\section{Conclusions}

This paper presented the final target selection of the 4MOST-Cosmological Redshift Survey (CRS) for the Bright Galaxy (BG) and Luminous Red Galaxy (LRG) samples.
A former selection was presented in~\citep{2019Msngr} based on the VISTA photometric survey composed of VHS~ \citep{vhs_dr5} and VIKINGS~\citep{viking_2013}. In this work, we have transitioned from the VISTA photometric survey to the Legacy Survey DR10.1, which combines DECaLS and DES data, as the photometric basis for 4MOST-CRS target selection. For LRGs, we continue to use WISE data but now combine it with Legacy Survey $g$, $r$, and $z$ bands instead of $J$ and $K_s$ bands from VISTA. This change is motivated by two main factors: (1) alignment with DESI-like colour cuts facilitates the combination of both surveys across the southern hemisphere, and (2) the photometry from the Legacy Survey is optimised for spectroscopic applications, enabling improved clustering measurements and better control of systematics in overdensity estimates. The ELG sample originally planned in the VISTA selection~\citep{2019Msngr} has been removed due to the limitation of fibre hours available for the CRS survey. Moreover, Euclid will cover a large part of the ELG targets in the 4MOST-CRS footprint.

\cref{sec:LS_target_sel} presents the new selection strategy. We have defined new colour cuts, closely following DESI’s, with slight modifications to remain within 4MOST-CRS fibre allocation limits while preserving the desired redshift distribution. Target selection tests using DESI DR1 data show low stellar contamination $\sim1.5\%$ for BGs and 0.5 to 4\% for LRGs. OpSim simulations (\cref{sec:comp}) predict survey completeness levels of 87\% for BGs and 67\% for LRGs after five years, reaching a Figure of Merit of 0.96 and 0.5, respectively. Additionally, \cref{sec:forecasts} presents Fisher forecasts which indicate that combining 4MOST-CRS samples with the DESI dataset using a 4MOST-like selection improves the constraints of $\sim 12\%$ on BAO and $\sigma_8$ in the redshift range $0.4 < z < 0.8$ compared to DESI alone. Finally, in \cref{sec:phot_sys} we applied linear and Random Forest regression methods to mitigate systematic effects in the DECaLS and DES photometric data, ensuring robust cosmological measurements from 4MOST-CRS.

The 4MOST-CRS will deliver the most extensive spectroscopic redshift catalogue in the southern hemisphere, achieving target densities of approximately 250 deg$^{-2}$ for BGs and 400 deg$^{-2}$ for LRGs over a 5700 deg$^2$ footprint. The 4MOST instrument has now arrived at the Paranal Observatory, where it is currently being installed on the VISTA telescope. 4MOST operations are expected to start later this year. This dataset will be a key resource for cross-correlation studies with upcoming surveys such as Vera Rubin Observatory LSST~\citep{LSST}, \textit{Euclid}~\citep{euclid}, HIRAX~\citep{HIRAX} or SKAO~\cite{SKA}.

\section*{Acknowledgements}
The authors thanks Andrei Variu and Anand Raichoor for the useful discussions and their contribution to the early stage of this work. AV, AR, JPK and MG acknowledge support from the Swiss National Science Foundation (SNF) “Cosmology with 3D Maps of the Universe” research grant 200020\_207379. BR acknowledges support from MNiSW grant DIR/WK/2018/12 and from grant pl0201-01 at the Pozna\'n Supercomputing and Networking Center. ET acknowledges funding from the HTM (grant TK202), ETAg (grant PRG1006) and the EU Horizon Europe (EXCOSM, grant No. 101159513). This work used the DiRAC Data Intensive service (CSD3) at the University of Cambridge, managed by the University of Cambridge University Information Services on behalf of the STFC DiRAC HPC Facility \footnote{\url{https://www.dirac.ac.uk}}. The DiRAC component of CSD3 at Cambridge was funded by BEIS, UKRI and STFC capital funding and STFC operations grants. DiRAC is part of the UKRI Digital Research Infrastructure. MB is supported by the Polish National Science Centre through grants no.\/ 2020/38/E/ST9/00395 and 2020/39/B/ST9/03494.
\\
The distribution of the work presented in this paper is as follows:
\begin{itemize}
    \item Aurélien Verdier: LRG target selection and forecasts
    \item Antoine Rocher: BG selection and photometric systematics
    \item Behnood Bandi: Masks and angular clustering analysis
    \item Boudewijn Roukema: Contributions to the software and to the preparation of the paper
    \item Elmo Tempel: Significant contribution to the 4MOST simulator (OpSim) and review of the paper
    \item Jon Loveday, Maciej Bilicki, Jean-Paul Kneib and Mathilde Guitton: Review of the paper
\end{itemize}





\bibliographystyle{mnras}
\bibliography{sample} 




\appendix

\section{DECaLS and DES target density correction}
\label{app-DECaLS-DES}
\begin{figure*}
    \centering
    \includegraphics[width=\linewidth]{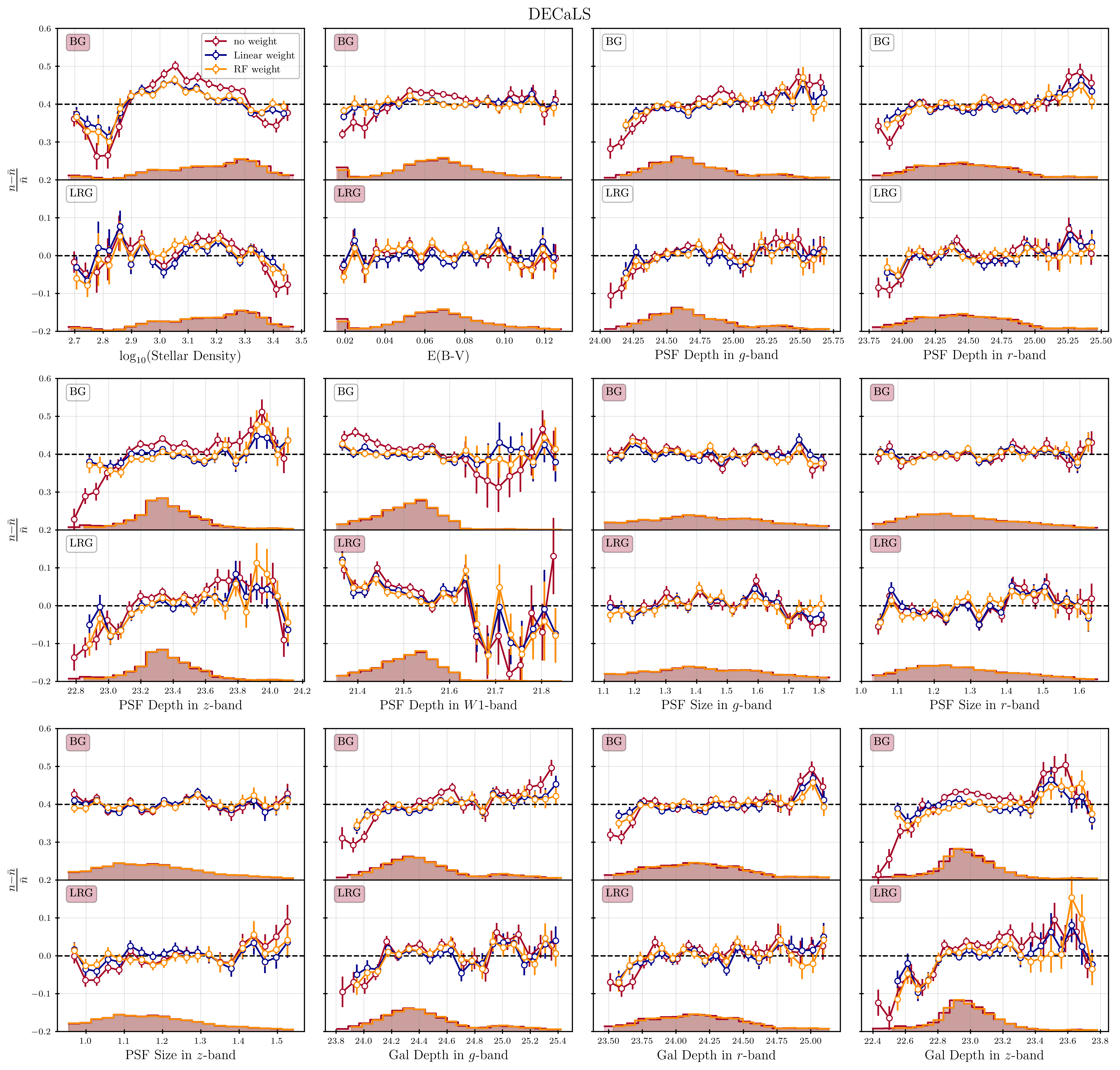}
    \caption{Relative target density as a function of each observational feature for the DECalS region. The red points correspond to the uncorrected (raw) density while the yellow and blue line correspond to the corrected density using linear and RF regressions. The y-axis values for the BGs sample has been shifted $+0.3$. The histogram represents the fraction of objects in each bin for each observational feature and the error bars are the estimated standard deviation of the normalized target density in each bin. The label “BG” or “LRG” is coloured by a red-shaded box when the feature was used for the regression, while a white box indicates that the features was not used for the regression.}
    \label{fig:DECaLS_sys}
\end{figure*}

\begin{figure*}
    \centering
    \includegraphics[width=\linewidth]{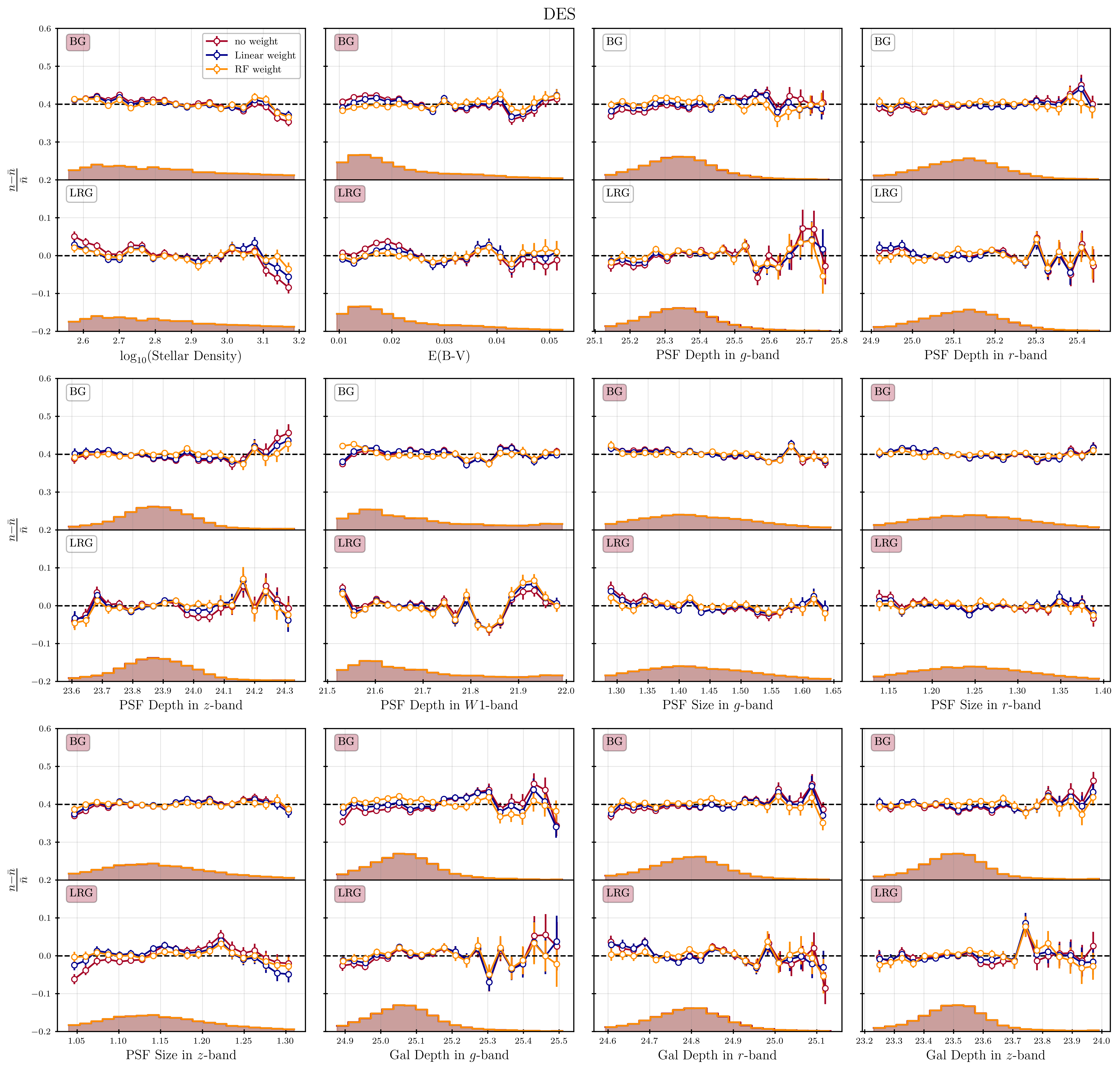}
    \caption{Same plot as \cref{fig:DECaLS_sys} for DES region.}
    \label{fig:DES_sys}
\end{figure*}

Results for DECaLS and DES mitigation using linear regression and Random Forest methods are respectively shown in \cref{fig:DECaLS_sys} and \cref{fig:DES_sys}.
Different parameters to measure systematic errors (see \cref{sec:sys_err}) are shown but the parameters used for mitigation are highlighted with a red label (instead of white) on the top left of each parameter box in the plots. 
For BGs, we use stellar density, $E(B-V)$, PSF size, and galaxy depth in the $g$, $r$, and $z$ bands.
For LRGs, the same parameters are selected except for stellar density, and the $W1$ band is added for the PSF size and galaxy depth.
The PSF Depth is not used in the bands where the galaxy depth is available, as both provide the same kind of information.
Overall, both methods correct the raw data overdensities well, especially for reducing the impact of the stellar density for DECaLS, where Random Forest shows better improvements than linear regression. For DES, there is not much improvement due to the fact that DES is already of good quality.

\bsp    
\label{lastpage}
\end{document}